\documentclass[12pt,preprint]{aastex}
\usepackage{epsf}

\begin{document} 

\title{The Size Distribution of Kuiper Belt Objects}

\author{Scott J. Kenyon}
\affil{Smithsonian Astrophysical Observatory, 60 Garden Street, 
Cambridge, MA 02138, USA; e-mail: skenyon@cfa.harvard.edu}
\author{and}
\author{Benjamin C. Bromley}
\affil{Department of Physics, University of Utah, 201 JFB, Salt Lake City, 
UT 84112, USA; e-mail: bromley@physics.utah.edu}

\begin{abstract}
We describe analytical and numerical collisional evolution 
calculations for the size 
distribution of icy bodies in the Kuiper Belt.  For a wide range 
of bulk properties, initial masses, and orbital parameters,
our results yield power-law cumulative size distributions, 
$N_C \propto r^{-q}$, with $q_L \approx$ 3.5 for large bodies 
with radii, $r \gtrsim$ 10--100 km, and 
$q_s \approx$ 2.5--3 for small bodies 
with radii, $r \lesssim$ 0.1--1 km.
The transition between the two power laws occurs at a break
radius, $r_b \approx$ 1--30 km.
The break radius is more sensitive to the initial mass in the 
Kuiper Belt and the amount of stirring by Neptune than the
bulk properties of individual Kuiper Belt objects (KBOs).
Comparisons with observations indicate that most models can 
explain the observed sky surface density $\sigma(m)$ of KBOs for 
red magnitudes $R \approx$ 22--27. For $R \lesssim$ 22 and 
$R \gtrsim$ 28, the model $\sigma(m)$ is sensitive to the amount 
of stirring by Neptune, suggesting that the size distribution 
of icy planets in the outer solar system provides independent 
constraints on the formation of Neptune.

\end{abstract}

\subjectheadings{planetary systems -- solar system: formation -- 
stars: formation -- circumstellar matter}

\section{INTRODUCTION}

The Kuiper Belt is a vast swarm of icy bodies beyond the orbit 
of Neptune in our solar system.  Following the discovery of the
first Kuiper Belt objects (KBOs) in 1930 \citep[Pluto;][]{tom46}
and 1992 \citep[1992 QB$_1$;][]{jew93}, several groups began 
large-scale surveys to characterize the limits of the Kuiper Belt
\citep[e.g.,][and references therein]{luu97,all01,gla01,lar01}.
Today, there are 800--1000 known KBOs with radii $r \gtrsim$ 
50 km in orbits that extend from 35 AU out to at least 150 AU 
\citep{luu02, ber03}. The vertical scale height of the KBO
population is $\sim$ 20--30 deg. The total mass is $\sim$ 
0.01--0.1 $M_{\oplus}$ \citep{luu02}.

Observations place numerous constraints on the apparent size
distribution of KBOs.  Deep imaging surveys at $R \approx$ 
22--28 suggest a power-law cumulative size distribution, 
$N_C \propto r^{-q_o}$, with $q_o = 3.0 \pm 0.5$ \citep{tru01, luu02}.  
Data from the {\it ACS} on {\it HST} suggest a change in the
slope of the size distribution at a break radius $r_b \sim$ 
10--30 km \citep{ber03}.
Dynamical considerations derived from the orbits of Pluto-Charon
and Jupiter family comets suggest $r_b \sim$ 1--10 km 
\citep{dun88,lev94,lev95,dun95,dun97,ip97,ste03}.  Observations of
the optical and far-infrared background light require
$q_o \lesssim$ 2.5 for small objects with radii $\lesssim$ 0.1--1 km
\citep{bac95, ste96b, tep99, kw01}, suggesting $r_b \gtrsim$ 0.1--1 km.

These observations provide interesting tests of planet formation 
theories.  In the planetesimal hypothesis, planetesimals with radii 
$\lesssim$ 1--10 km collide, merge, and grow into larger objects.  
This accretion process yields a power-law size distribution, 
$N_C \propto r^{-q}$, with $q_L \approx$ 2.8--3.5 
for 10--100 km and larger objects \citep{kl99b,ken02}.  
As the largest objects grow, their gravity stirs up the orbits of 
leftover planetesimals to the disruption velocity. Collisions
between leftover planetesimals then produce fragments instead
of mergers.  For objects with radii of 1--10 km and smaller,
this process yields a power-law size distribution with a 
shallower slope, $q_S \approx$ 2.5
\citep{ste96a, dav97, st97a, st97b, kl99a, kb04}.  

Despite uncertainties in the observations and the theoretical
calculations, the predicted power-law slopes for large and small
KBOs agree remarkably well with the data.  However, many issues
remain.  The observed sky surface density of the largest objects
with radii of 300--1000 km
and the location of the break in the size distribution are
uncertain \citep[see, for example,][]{luu02,ber03}. The data 
also suggest that different dynamical classes of KBOs have 
different size distributions \citep{lev01,ber03}.  
Theoretical models for the formation of KBOs
have not addressed this issue. Theoretical predictions for 
$r_b$, $q_S$, and the space density of KBOs are also uncertain. 
Observations indicate that the observed space density of KBOs is 
$f \lesssim$ 1\% of the initial density of solid material in the 
planetesimal disk.  Theoretical estimates of long-term collisional 
evolution yield $f \lesssim$ 10\% \citep[e.g.,][]{st97b,ken02}, but 
the sensitivity of this estimate to the bulk properties of KBOs 
has not been explored in detail.  

Here, we consider collision models for the formation and long-term
collisional evolution of KBOs in the outer solar system.  We use 
an analytic model to show how the break radius depends on the bulk 
properties and orbital parameters of KBOs 
\citep[see][for a similar analytic model]{pan04} and confirm 
these estimates with numerical calculations.  If KBOs have 
relatively small tensile strengths and formed in a relatively 
massive solar nebula, we derive a break radius, $r_b \sim $ 
3--30 km, close to the observational limits. The numerical 
calculations also provide direct comparisons with the observed
sky surface density and the total mass in the Kuiper Belt. Models
with relatively weak KBOs and additional stirring by Neptune
yield the best agreement with observations and make testable
predictions for the surface density of KBOs at $R \approx$ 
28--32.  In the next 3--5 yr, occultation observations can 
plausibly test these predictions.

We develop the analytic model in \S 2, describe numerical
simulations of KBO evolution in \S 3, and conclude with a 
brief discussion and summary in \S 4.

\section{ANALYTIC MODEL}

\subsection{Derivation}

We begin with an analytic collision model for the long-term
evolution of an ensemble of KBOs.  We assume that KBOs lie
in an annulus of width $\Delta a$ centered at a heliocentric 
distance $a$. KBOs with radius $r$ orbit the Sun with eccentricity 
$e$ and inclination $i$.  The scale height $H$ of the KBO population
is $H$ = $a$ sin $i$. 

KBOs evolve through collisions and long range gravitational interactions. 
Here, we assume that gravitational interactions have reached a 
steady-state, with $e$ and $i$ constant in time.  Based on numerical
simulations of KBO formation, we adopt a broken power law for the 
initial size distribution, 

\begin{equation}
n(r) = \left\{ \begin{array}{l l l}
	n_S r^{-\alpha_S} & \hspace{5mm} & r \le 1 ~ {\rm km} \\
\\
	n_L r^{-\alpha_L} & \hspace{5mm} & r \ge 1 ~ {\rm km} \\
         \end{array}
         \right .
\end{equation}

\noindent
with $\alpha_S$ = 3.5 for small objects and $\alpha_L$ = 4.0 for
large objects \citep{st97a, dav97, st97b, kl99a, ken02, kb04}.
If $V$ is the relative velocity, the collision
rate for a KBO with radius $r_1$ and all KBOs with radius $r_2$ is
\begin{equation}
\frac{dn_1}{dt} = \left ( \frac{n_2}{4~H~a~\Delta a} \right ) ~
\left ( 1 + 2.7 \frac{V_e^2}{V^2} \right ) ~
V~ (r_1 + r_2)^2 ~ ,
\end{equation}
where $V_e$ is the escape velocity of a single body with mass, 
$m = m_1 + m_2$ \citep[][and references therein]{ws93, kl98}.

The collision outcome depends on the impact velocity $V_I$ and
the disruption energy $Q_d$. We follow previous investigators
and define the energy needed to remove 50\% of the combined mass
of two colliding planetesimals,
\begin{equation}
Q_d = Q_b r^{\beta_b} + \rho Q_g r^{\beta_g} ~ ,
\end{equation}
where $Q_b r^{\beta_b}$ is the bulk (tensile) 
component of the binding energy and 
$\rho Q_g r^{\beta_g}$ is the gravity component 
of the binding energy 
\citep[see, for example,][]{dav85,ws93,hls94,ben99,hou99}. 
The gravity component of the disruption energy varies linearly 
with the mass density $\rho$ of the planetesimals.
This expression ignores a weak relation between $Q_b$,$Q_g$
and the impact velocity $V_I$
\begin{equation}
Q_b, Q_g \propto \left ( \frac{V_I}{V_0} \right ) ^{\beta_v} ~ ,
\end{equation}
where $\beta_v \approx$ 0.25--0.50 for rocky material
and $\beta_v \approx$ $-0.25$ to $-0.50$ for icy material
\citep[e.g.][]{hou90,hou99,ben99}. For an analytic model with
$V_I$ = constant, the variation of $Q_d$ with $V_I$ is
not important. We consider non-zero $\beta_v$ in complete 
evolutionary calculations described below.

We adopt the standard center of mass collision energy
\citep{ws93}
\begin{equation}
Q_I = \frac{m_1 m_2 V_I^2}{4 (m_1 + m_2)^2} ~ ,
\end{equation}
where the impact velocity is
\begin{equation}
V_I^2 = V^2 + V_e^2 ~ .
\end{equation}
The mass ejected in a collision is
\begin{equation}
m_{ej} = 0.5 (m_1 + m_2) \left ( \frac{Q_I}{Q_d} \right )^{\beta_e} ~ ,
\end{equation}
where $\beta_e$ is a constant of order unity 
\citep[see][and references therein]{dav85, ws93, kl98, ben99}.

For a single KBO, the amount of mass accreted in collisions with all 
other KBOs during a time interval $\delta t$ is
\begin{equation}
\delta m_a (r) = \dot{m_a} ~ \delta t = \delta t ~ \int \frac{dn_1}{dt} ~ m_2 ~ dm_2 ~ .
\end{equation}
The amount of mass lost is
\begin{equation}
\delta m_l (r) = \dot{m_l} ~ \delta t = \delta t ~ \int \frac{dn_1}{dt} ~ m_{ej} ~ dm_2 ~ .
\end{equation}
KBOs with $\dot{m_l} > \dot{m_a}$ lose mass and reach zero 
mass on a removal timescale
\begin{equation}
t_r(r) \approx \frac{m_0(r)}{\dot{m_l} - \dot{m_a}}  ~ .
\end{equation}

With these definitions, the evolution of an ensemble of KBOs
depends on the relative velocity $V$, the size distribution,
and the disruption energy.  Because the disruption energy
scales with size, larger objects are harder to disrupt than 
smaller objects. To produce a break in the size distribution, 
we need a `break radius,' $r_b$, where 
\begin{equation}
\begin{array}{l l l}
	t_r \le t_0 & \hspace{5mm} & {\rm for} ~ r \le r_b \\
\\
	t_r > t_0 & \hspace{5mm} & {\rm for} ~ r > r_b \\
         \end{array}
         .
\end{equation}
and $t_0$ is some reference time. We choose $t_0$ = 1 Gyr as
a reasonable e-folding time for the decline in the KBO space 
density.

\subsection{Application}

To apply the analytic model to the KBO size distribution, we use
parameters appropriate for the outer solar system. We adopt $i$ = 
$e$/2, with $e$ = 0.04 for classical KBOs and $e$ = 0.2 for Plutinos.
This simplification ignores the richness of KBO orbits, but gives 
representative results without extra parameters.
We assume a total mass in KBOs, $M_{KBO}$, in an annulus with 
$a_0$ = 40 AU and $\delta a$ = 10 AU.  A minimum mass solar
nebula has $M_{KBO} \sim$ 10 $M_{\oplus}$; the current Kuiper Belt 
has $M_{KBO}$ = 0.05--0.20 $M_{\oplus}$ 
\citep{ste96a,luu02,ber03}.  
This range in initial KBO mass provides a representative range for 
the normalization constants, $n_S$ and $n_L$, in our model size 
distribution.

To model the destruction of KBOs, we adopt representative values
for $Q_b$, $Q_g$, $\beta_b$, $\beta_g$, and $\rho$. Because the
bulk properties of KBOs are poorly known, we consider wide
ranges in $Q_b = 10^1$ to $10^8$ erg g $^{-1}$, 
$C_g = \rho Q_g$ = $10^{-4}$ to $10^4$ erg g$^{-1}$, 
and $\beta_g$ = 0.5--2.0. These ranges span analytic and numerical results 
in the literature \citep[e.g.,][]{dav85,hls94,lov96,ben99,hou99}. 
For simplicity, we adopt $\beta_b$ = 0; other choices have 
little impact on the results.

Figure 1 illustrates several choices for the disruption energy.
We adopt a normalization constant for the gravity component of
the disruption energy,
\begin{equation}
C_g = \rho ~ Q_g = C_0 ~ \rho ~ (10^5)^{1.25 - \beta_g} ~ ,
\end{equation}
with $\rho$ = 1.5 and $C_0$ = 1.5.
All model curves then have the same gravity component of $Q_d$ 
at $r$ = 1 km.  The horizontal lines plot the impact energy for two
massless KBOs with $e$ = 0.001 (lower line), 0.01 (middle line),
and 0.1 (upper line). When $e \le$ 0.01, collisions cannot disrupt 
large KBOs with $r \gtrsim$ 1 km. Collisions disrupt 
smaller KBOs only if $Q_b \lesssim 10^6$ erg g$^{-1}$ (e/0.01)$^2$.  
When $e \sim$ 0.1, collisions disrupt all small KBOs independent 
of their bulk strength. For KBOs with $r \ge$ 1 km, disruptions are 
sensitive to $\beta_g$ and $C_g$. In general, more compact KBOs
with larger $\beta_g$ and $C_g$ are harder to disrupt than 
fluffy KBOs with small $C_g$.

Figure 2 plots several realizations of the analytic model
as a function of KBO radius for a model with a mass equal
to the minimum mass solar nebula and fragmentation parameters
listed in the legend.
For any combination of $Q_b$ and $\beta_g$, the removal 
timescale $t_r$ is a strong function of the KBO radius.  
Small KBOs with $r \le$ 1 km have large $\dot{m_l}/m_0$ and 
small removal timescales of 1 Myr or less. Larger objects 
lose a smaller fraction of their initial mass per unit time 
and have $t_r \sim$ 1--1000 Gyr.  KBOs with smaller $\beta_g$ 
are more easily disrupted and have smaller disruption times 
than KBOs with larger $\beta_g$.

Figure 2 also illustrates the derivation of the break radius, $r_b$.
The horizontal line at $t$ = 1 Gyr intersects the model curves
at $r_b \sim$ 3 km ($\beta_g$ = 2) and at $r_b \sim$ 6 km 
($\beta_g$ = 1.25).  The break radius is clearly independent
of the bulk strength $Q_b$, and the reference time $t_0$,
but is sensitive to $\beta_g$ and $e$. At fixed collision 
energy, KBOs with smaller $\beta_g$ 
fragment more easily. Thus, the break radius becomes larger 
as $\beta_g$ becomes smaller. At fixed strength, KBOs with 
larger impact velocities also fragment more easily.  Thus, 
the break radius becomes larger with larger $e$.

As in \citet{pan04}, the break radius depends on the initial
mass in KBOs and the collision velocity. For classical KBOs
with $e \approx$ 0.04, the break radius is $r_b \lesssim$
10--20 km when $M_{KBO} \lesssim$ 10 $M_{\oplus}$ (Figure 3).
In our model, $r_b$ grows with the initial mass in KBOs.
We derive roughly an order of magnitude change in $r_b$
for a two order of magnitude change in the initial mass in
KBOs.

The break radius is also sensitive to $C_g$ and $\beta_g$.
KBOs with small $C_g$ and $\beta_g$ are easier to fragment
than KBOs with large $C_g$ and $\beta_g$. For a fixed mass
in KBOs, plausible variations in $C_g$ and $\beta_g$ yield
order of magnitude variations in $r_b$ (Figure 3). At large
$Q_b$, these differences disappear. For a Kuiper Belt with 1\%
of the mass in a minimum mass solar nebula, the model predicts
no dependence of $r_b$ on the physical variables for
$Q_b \gtrsim 3 \times 10^6$ erg g$^{-1}$. This limit lies at $Q_b$
$\gtrsim 10^8$ erg g$^{-1}$ for a minimum mass solar nebula.

Figure 4 shows how changes in the collision energy modify $r_b$.
Models with large eccentricity, $e$ = 0.20, yield $r_b \approx$ 
3--100 km, compared to 0.3--10 km for $e$ = 0.04.  The predicted
$r_b$ is sensitive to the initial mass in KBOs, but is less
sensitive to $C_g$ and $\beta_g$.

In contrast to \citet{pan04}, our analytic model indicates that 
a large break radius does not require a small bulk strength for 
the planetesimals. When the initial mass in KBOs is small,
$\sim$ 1\% to 10\% of the minimum mass solar nebula, the
collision frequency for 10--100 km KBOs is also small, $\sim$ 
10--100 Gyr$^{-1}$. To remove sufficient material in 5 Gyr, 
these collisions must produce mostly debris. Thus, $Q_d$ must be 
small. Large bulk strengths, $Q_b \gtrsim 10^6$ erg g$^{-1}$, 
preclude debris-producing collisions (Figure 1) and result in
small $r_b$ (Figures 3--4).  As the initial mass in KBOs increases
to 10\% to 100\% of the minimum mass solar nebula, collision 
frequencies also increase. More frequent collisions can remove
large KBOs from the size distribution even when $Q_b$ is large.
Thus, the break radius is less sensitive to $Q_b$ for massive
nebulae. 

\subsection{Implications for the Kuiper Belt}

The conclusions derived from the analytic model have direct 
implications for the formation and evolution of KBOs.  The
apparent break in the sky surface density at $R \approx$ 28
requires a break in the size distribution at $r_b \approx$
20 km for an albedo of 0.04.  The analytic model shows that
a measured
$r_b \gtrsim$ 20 km requires a nebula with a mass in solids 
of at least 10\% of the minimum mass solar nebula.  Producing
such a large break radius from collisions is easier in an
initially more massive nebula.  The
current mass in KBOs is $\lesssim$ 1\% of the minimum mass
solar nebula.  Thus, the analytic model provides additional 
support for an initially more massive solar nebula at 30--50 AU
\citep[see also][and references therein]{st97a, kl99a, ken02}.  

The break in the sky surface density also favors a low bulk 
strength, $Q_b \lesssim 10^6$ erg g$^{-1}$, for KBOs. This
$Q_b$ is smaller than the $Q_b \approx 10^7$ erg g$^{-1}$ 
derived from numerical models of collisions of icy bodies
\citep[e.g.,][]{ben99}. However, a low bulk strength is 
consistent with the need for a strengthless rubble pile
in models of the break-up of comet Shoemaker-Levy 9
\citep[e.g.,][]{asp96}. 

The analytic model may also explain differences in the observed 
size distributions for different dynamical classes of KBOs. 
From Figures 3--4, resonant KBOs and scattered KBOs with large 
$e$ and $i$ should have a larger $r_b$ than classical KBOs with 
small $e$ but large $i$ \citep{luu02}. The group of `cold' 
classical KBOs with small $e$ and small $i$ \citep{lev01}
should have even smaller $r_b$. The observations provide
some support for this division. Relative to the number of 
bright KBOs, there are fewer faint resonant and scattered 
KBOs and more classical KBOs than expected \citep{ber03}.  
Long-term collisional evolution could be responsible for
removing higher velocity resonant and scattered KBOs and
leaving behind lower velocity classical KBOs.

To test the analytic model and provide direct comparisons with
the observations, we need a numerical model of accretion and
erosion.  A numerical model accurately accounts for the time
dependence of the total mass in KBOs and thus provides a clear 
measure of the removal time for a range of sizes. Numerical models 
also yield a direct calculation of the size distribution and 
thus measure the `depletion' of KBOs as a function of radius. 

\section{NUMERICAL MODELS}

To test the analytic model, we examine numerical simulations with a 
multiannulus coagulation code \citep[][and references therein]{kb04}.
For a set of $N$ concentric annuli surrounding a star of mass $M$,
this code solves numerically the coagulation and Fokker-Planck equations 
for bodies undergoing inelastic collisions, drag forces, and long-range 
gravitational interactions \citep{kb02}.  We adopt collision rates 
from kinetic theory and use an energy-scaling algorithm to assign 
collision outcomes \citep{dav85,ws93,wei97,ben99}.  We derive changes 
in orbital parameters from gas drag, dynamical friction, and viscous 
stirring \citep{ada76, oht02}.  The appendix describes updates to
algorithms described in \citet[][2002a, 2004]{kb01} and \citet[1999]{kl98}.

We consider two sets of calculations. For models without gravitational
stirring, we set $e$ = constant, $i$ = $e$/2, and calculate the
collisional evolution of an initial power law size distribution.
These calculations require a relatively small amount of computer 
time and allow a simple test of the analytic model. 

Complete evolutionary calculations with collisional evolution
and gravitational stirring test whether particular outcomes 
are physically realizable. These models also provide direct
tests with observables, such as the current mass in KBOs and
the complete KBO size distribution. Because these models do
not allow arbitrary $e$ and $i$, they are less flexible than
the constant $e$ models.  

To provide some flexibility in models with gravitational stirring, 
we calculated models with and without stirring by Neptune at 30 AU.
In models without Neptune, large KBOs with radii of 1000--3000 km
stir up smaller KBOs to the disruption velocity. The KBO size
distribution, including the break radius, then depends on the
radius of the largest KBO $r_{L,KBO}$ formed during the calculation. 
Because $r_{L,KBO}$ depends on $Q_b$ and $Q_g$ \citep[e.g.,][]{kl99a, ken02},
$r_b$ also depends on $Q_b$ and $Q_g$.

In models with Neptune, long-range stirring by Neptune can dominate
stirring by local large KBOs.  The break radius then depends on
the long-range stirring formula \citep{wei89,oht02} and the timescale
for Neptune formation. Here, we assume a 100 Myr formation time
for Neptune, whose semimajor axis is fixed at 30 AU throughout the
calculation. The mass of Neptune grows with time as 
\begin{equation}
M_{Nep} \approx \left\{ \begin{array}{l l l}
	6 \times 10^{27} ~ e^{(t-t_0)/t_1} ~ {\rm g} & \hspace{5mm} & t < t_0 \\
\\
	6 \times 10^{27} ~ {\rm g} ~ + ~ C_{Nep}(t-t_1) & \hspace{5mm} & t_0 < t < t_2 \\
\\
	1.0335 \times 10^{29} ~ {\rm g} & \hspace{5mm} & t > t_2 \\
         \end{array}
         \right .
\end{equation}
where $C_{Nep}$ is a constant and $t_0$, $t_1$, and $t_2$ are
reference times. For most calculations, we set $t_0$ = 50 Myr,
$t_1$ = 3 Myr, and $t_2$ = 100 Myr.  These choices allow our
model Neptune to reach 1 $M_{\oplus}$ in 50 Myr, when the 
largest KBOs have formed at 40--50 AU, and reach its current 
mass in 100 Myr. This prescription is not intended as a model
for Neptune formation, but it provides sufficient extra stirring
to test the prediction that the break radius depends on the
amount of local stirring.

\subsection{Constant eccentricity models}

Calculations with constant eccentricity allow a direct test of the 
analytic model. We performed a suite of $\sim$ 200 4.5 Gyr calculations 
for a range in fragmentation parameters, with log $Q_b$ = 1--8, 
$\beta_g$ = 0.5--2.0, and log $C_g$ + 5 $\beta_g$ = 0.01--20.
The initial size distribution of icy planetesimals
has sizes of 1 m to 100 km in mass bins with $\delta = m_{i+1}/m_i$
= 1.7 and equal mass per mass bin. The planetesimals lie in 32 annuli
extending from 40~AU to 75~AU.  The central star has a mass of 
1 M$_{\odot}$. The initial surface density, 
$\Sigma_0 = 10^{-3}$ g~cm$^{-2}$ to $10^{-1}$ g~cm$^{-2}$ at 40 AU, 
ranges from 1\% to 100\% of the minimum mass solar nebula extended 
to the Kuiper Belt \citep{wei77,hay81}.  The initial eccentricity, 
$e$ = 0.04 and 0.20, spans the observed range for classical and 
resonant KBOs \citep{luu02}.  

In Kuiper Belt models with large $e$, fragmentation is the dominant
physical process \citep[see also][]{st97a, kl99a, kb02}. Large, 
10--100~km
objects grow very slowly.  Smaller objects suffer numerous disruptive 
collisions that produce copious amounts of debris. Debris fills lower 
mass bins, which suffer more disruptive collisions.  In 10--100 Myr, 
this collisional cascade reduces the population of 0.1--1~km and 
smaller objects. The size distribution then follows a broken power 
law, with $\alpha_L \approx$ 3 for large objects and 
$\alpha_S \approx$ 2.5 for the small objects \citep{doh69,wil94,pan04}.

As the evolution proceeds, the size distribution evolves into a standard 
shape (Figure 5). After 4.5 Gyr, the largest objects grow from 
100 km to 125 km. 
From $\sim$ 10 km to 125 km, the size distribution continues to 
follow a power-law with $\alpha_L \approx$ 3. Test calculations 
suggest that this power law slope is fairly independent of the
initial power law.

At smaller sizes, the shape of the size distribution depends on 
the bulk strength.
For $Q_b \gtrsim 10^6$ erg g$^{-1}$, disruptive collisions produce 
a break in the size distribution; the power law slope changes
from $\alpha_L \approx$ 3 to $\alpha_S \approx$ 2.5. For 
$Q_b \lesssim 10^4$ erg g$^{-1}$, disruptive collisions produce 
two breaks, one at 1--10 km and another at $\sim$ 0.1 km (Figure 5).  
The break at 1--10 km is where growth by accretion roughly balances 
loss by disruption.  The break at $\sim$ 0.1 km is where debris 
produced by collisions of larger objects roughly balances loss 
by disruptive collisions. Between these two sizes, the slope of 
the size distribution ranges from
$\alpha_I \approx -0.5$ for $Q_b \sim 10$ erg g$^{-1}$ to
$\alpha_I \approx 1$ for $Q_b \sim 10^5$ erg g$^{-1}$. The slope
of the power law for small sizes ranges from 
$\alpha_I \approx$ 4.5--5.0 for $Q_b \sim 10$ erg g$^{-1}$ to
$\alpha_I \approx$ 3 for $Q_b \sim 10^5$ erg g$^{-1}$. 
As $Q_b$ approaches $10^6$ erg g$^{-1}$, the slopes of both 
power laws converge to $\alpha \approx$ 2.5.

We define the first inflection point in the size distribution as
the break radius. To measure $r_b$, we use a least-squares fit to
derive the best-fitting power-law slopes, $\alpha_L$ and $\alpha_I$,
to the calculated size distribution. Using Poisson statistics to
estimate errors in $N$, we derive $r_b$ and its 1$\sigma$ error by 
minimizing the residuals in the fits. Typical errors in log $r_b$
are $\pm$ 0.02-0.05. Tests indicate that the derived $r_b$ is more
sensitive to the mass resolution of our calculations, $\delta$,
than to the range in log $r$ used in the fits. This error is also small
compared to the range in log $r_b$, $\sim$ 0.2, derived from repeat 
calculations with the same combination of $Q_b$ and $\beta_g$.

Figure 6 shows results for calculations with constant $e$ = 0.04.
For all initial masses and $Q_b \lesssim 10^6$, $r_b$ is independent 
of $Q_b$. For small initial masses, calculations with stronger objects 
yield small break radii. At larger initial masses, this sensitivity 
to $Q_b$ disappears. Calculations for minimum mass solar nebulae show
little variation of $r_b$ with $Q_b$.

These results confirm the basic features of the analytic model. 
Both models predict $r_b \lesssim$ 1 km for low mass nebulae
with $\sim$ 1\% of the mass in the minimum mass solar nebula. 
Larger break radii, $\sim$ 1--10 km, are possible in more
massive nebulae. The analytic model predicts large break radii
for larger $e$ than the numerical calculations. In numerical 
calculations with $e$ = 0.2, disruptive collisions reduce the 
space density considerably in $\sim$ 100 Myr. The smaller 
collision rates prevent formation of a break in the size distribution
at large radii.  Thus, numerical calculations with $e$ = 0.2 yield 
log $r_b$ only $\sim$ 0.1--0.2 larger than calculations with 
$e$ = 0.04.

\subsection{Full evolution models}

These calculations begin with 1--1000 m planetesimals in mass bins 
with $\delta$ = 1.4 or 1.7 and equal mass per bin. The planetesimals
lie in 32 annuli at 40--75 AU. Models with Neptune have an extra
annulus at 30 AU. For most models, we adopt $e_0 = 10^{-4}$ or
$e_0$ = $10^{-5}$ and $i$ = $e$/2 for all planetesimals.  At the 
start of our calculations, these initial values yield a rough balance
between viscous stirring by 0.1--1 km objects and collisional
damping of 10--100 m objects.  The bodies have a mass density 
$\rho_d$ = 1.5 g cm$^{-3}$, which is fixed throughout the evolution.  
We consider a range in initial surface density, with 
$\Sigma_0$ = 0.03--0.3 g cm$^{-2}$ ($a_0$/30 AU)$^{-3/2}$.  

To measure the sensitivity of our results to stochastic variations,
we performed 2--5 calculations for each set of fragmentation parameters.
For $\beta_b$ = 0 and a factor of ten range in $\Sigma_0$, we 
considered log $Q_b$ = 
1, 2, 3, 4, 5, 6, and 7; $C_0$ = 0.15 and 1.5; and
$\beta_g$ = 1.25 and 2.0.  We also performed a limited set of calculations 
for $C_g$ = 0.15 and 1.5, $\beta_g$ = 0.5, and a small set of
log $Q_b$. Although stochastic variations can change the size of 
the largest object at 40--50 AU, repeat calculations with identical
initial conditions yield small changes in the shape of the size 
distribution or the location of the break radius.  A few calculations
with $\beta_b \neq$ 0 yield interesting behavior in the size
distribution at 1--100 m sizes, but $r_b$ does not change dramatically.
A larger suite of calculations with $\beta_v \neq$ 0 leads to similar
conclusions.  We plan to report on these aspects of the calculations 
in a separate paper.

Icy planet formation in the outer solar system follows a standard 
pattern \citep[see][]{kl99a,kb04,gol04}. Small planetesimals
with $r_i \lesssim$ 1 km first grow slowly. Collisonal
damping brakes the smallest objects. Dynamical friction
brakes the largest objects and stirs up the smallest objects.
Gravitational focusing factors increase, and runaway growth
begins.  At 40--50 AU, it takes $\sim$ 1 Myr to produce 10 km 
objects and another 3--5 Myr to produce 100 km objects.  
Continued stirring reduces gravitational focusing factors.
Collisions between the leftover planetesimals produce debris
instead of mergers. Runaway growth ends and the collisional
cascade begins.  

During the collisional cascade, the mass in 1--10 km and smaller
objects declines precipitously.  Because gravitational focusing
factors are small, collisions between two planetesimals are more 
likely than collisions between a planetesimal and a 100--1000 km
protoplanet.  Thus, disruptive collisions grind leftover planetesimals
into small dust grains, which are removed by radiation pressure 
and Poynting-Robertson drag 
\citep[e.g.,][]{bur79,tak01}. 
At 40--50 AU, the surface density falls by a factor of two in
100--200 Myr, a factor of 4--5 in 1 Gyr, and more than an order
of magnitude in 3--4 Gyr \citep[see also][]{kb04}.  After 4.5 Gyr, 
the typical amount of 
solid material remaining at 40--50 AU is 3\% to 10\% of the
initial mass (see below).

As collisions and radiation remove material from the system, 
the largest objects continue to grow slowly. In most calculations, 
it takes 10--50 Myr to produce the first 1000 km object.  The 
largest objects then double their mass every 100 Myr to 1 Gyr. 
After 4.5 Gyr, the largest objects have radii ranging from $\sim$ 
100 km ($Q_b \lesssim 10^3$ erg g$^{-1}$, $\beta_g$ = 0.5)
to 5000 km ($Q_b \gtrsim$ $10^7$ erg g$^{-1}$, $\beta_g$ = 2.0).
Calculations with $\beta_g$ = 1.25 and $Q_b \sim 10^2$ erg g$^{-1}$
to $10^4$ erg g$^{-1}$ favor the production of objects with radii
of 1000--2000 km, as observed in the outer solar system.

These general results are remarkably independent of the initial
conditions and of some input parameters \citep{kl99a,kb04}.  The size 
distribution of objects remaining at 4.5 Gyr is not sensitive to 
the initial disk mass, the initial size distribution, the initial 
eccentricity and inclination (for $e_0 \lesssim 10^{-3}$), the mass 
resolution $\delta$, the width of an annulus $\delta a$, or the 
gas drag parameters.  The orbital period $P$ and the surface density 
set the collision timescale, $t \propto P / \Sigma $.  Although 
stochastic variations can produce factor of 2 or smaller variations
in growth times, all timescales depend on the collision time and
scale with the current surface density. 

To derive $r_b$ for these calculations, we again examine the size 
distribution at 4.5 Gyr (Figure 7). For radii $r_i \sim$ 10--1000 km, 
the size 
distribution follows a power-law, $N \propto r^{-\alpha_L}$ with slope 
$\alpha_L \approx$ 3--3.5. For smaller radii, the size distribution 
has inflection points at log $r_i \approx$ $-$1 to 1 and at 
log $r_i \lesssim$ $-$1.  Between the two inflection points, the 
power-law slope is shallow, with $\alpha_I \approx$ 0--2. For
small objects, the power-law slope is steep, with $\alpha_S \approx$ 2--4.

The slope of the intermediate power-law depends on the fragmentation
parameters.  Calculations with small $Q_b$ and $\beta_g$ yield small
$\alpha_I$. For larger $Q_b$ and $\beta_g$, the slope approaches the
collisional limit, $\alpha_I \approx$ 2.5 \citep{doh69, wil94}.  The 
extent of the shallow, intermediate power-law also varies with $Q_b$ 
and $\beta_g$. For $Q_b \lesssim 10^3$ erg g$^{-1}$ and $\beta_g \approx$
1.25, the size distribution is relatively flat from log $r_i \approx$
$-1$ to 0.0--0.5 (Figure 7).  Large $Q_b$ and $\beta_g$ result in a 
smaller extent for the intermediate power law.

Figure 8 shows results for $r_b$ as a function of log $Q_b$
for several sets of calculations. All calculations began with a
mass in solids comparable to a minimum mass solar nebula. For models
without stirring by Neptune, we derive
log $r_b \approx$ 5.0--5.6 (log $Q_b \lesssim$ 4) and
log $r_b \approx$ $5.3\pm0.3 - 0.5$ log $Q_b$ (log $Q_b \gtrsim$ 4).
Although there is some overlap in the results for different sets
of parameters, a weaker gravity component to the disruption energy 
($\beta_g \approx$ 1.25) favors larger $r_b$.  
This result confirms the conclusion derived from the analytic model.

Stirring by Neptune yields larger values for the break radius 
(Figure 8). As Neptune reaches its final mass at 80--100 Myr,
long-range stirring rapidly increases the eccentricities of
objects at 40--50 AU.  This stirring accelerates the collisional 
cascade, which depletes the population of small planetesimals 
and halts the growth of the largest objects.  The larger $e$ and 
longer duration of the collisional cascade moves the break in the 
size distribution to larger radii.

Figure 9 shows size distributions at 4.5 Gyr for three models
with stirring by Neptune.  The legend lists input values for
$Q_g$ and log $Q_b$.  In these calculations, the break is at
$r_b \approx$ 5--10 km, compared to $r_b \approx$ 1--5 km for
models without Neptune stirring. The position of the break 
is relatively independent of $Q_b$ or $Q_g$ (see Figure 8).  
For KBOs with sizes smaller than the break, the slope of the 
size distribution is sensitive to $Q_b$ but not to $Q_g$. 
Models with $Q_b \gtrsim 10^4$ 
erg g$^{-1}$ have more small objects with radii of $\sim$ 
0.1 km than models with $Q_b \lesssim 10^4$ erg g$^{-1}$.

\subsection{Comparisons with observations of KBOs}

The analytic model and the numerical evolution calculations 
yield a consistent picture for the size distribution of icy
bodies in the outer solar system.  The general shape of the size 
distribution does not depend on the initial conditions or
input parameters. The typical size distribution has two
power laws -- one power-law for large objects with radii
$\gtrsim$ 10--100 km and a second power-law for small objects
with radii $\lesssim$ 0.1 km -- connected by a transition
region where the number of objects per logarithmic mass
bin is roughly constant.  The power law slope for the large 
objects is also remarkably independent of input parameters 
and initial conditions.

The fragmentation parameters and the amount of stirring set the 
location of the transition region and the power-law slope for 
the small objects.  In our calculations, the break radius is
$r_b \gtrsim$ 10 km when icy objects are easy to break and 
the stirring is large.  Strong icy objects and small stirring
favor a small break radius, $r_b \lesssim$ 1 km. When the
break radius is small, the extent of the transition region 
is also small, less than an order of magnitude in radius.
When the break radius is large, the transition region can 
extend for 2 orders of magnitude in radius.

To compare the numerical results to observations of KBOs, we 
convert a calculated size distribution into a predicted
sky surface density of KBOs as a function of apparent 
magnitude. Current observations suggest that the observed sky
surface density, $\sigma (m)$ -- the number of KBOs per square
degree on the sky, follows a power law
\begin{equation}
{\rm log} ~  \sigma(m) = {\rm log} ~  \sigma_0  + \alpha (m - m_0)
\end{equation}
where $\alpha$, $\sigma_0$ and $m_0$ are constants. With 
$\alpha \approx$ 0.6 and $m_0$ = 23, this function fits the data 
fairly well for R-band magnitudes, $R \approx$ 22--26.  
For $R \lesssim$ 22 and $R \gtrsim$ 26, the simple function 
predicts too many KBOs compared to observations
\citep[][]{ber03}. 
The observations also suggest that the scattered and Plutino
populations of KBOs have different surface density distributions
than classical KBOs, with $\alpha \approx$ 0.6 for scattered
KBOs and Plutinos and $\alpha \approx$ 0.8 for classical KBOs
\citep{ber03}.

Because we do not include the dynamics of individual objects in our 
calculations, we cannot predict the relative numbers of KBOs in 
different dynamical classes. However, we can predict $\sigma(m)$
for all KBOs and see whether the calculations can explain trends 
in the observations.

To derive a model $\sigma(m)$, we assign distances $d_{\odot}$ to an 
ensemble of objects chosen randomly from the model size distribution. 
For a random phase angle $\beta$ between the line-of-sight from 
the Earth to the object and the line-of-sight from the Sun to 
the object, the distance of the object from the Earth is 
$d_E = d_{\odot} {\rm cos} \beta - (1 + d_{\odot}^2 ({\rm cos}^2 
\beta - 1))^{1/2}$. The red magnitude of this object is 
$R = R_0 + 2.5~{\rm log}~(t_1/t_2) - 5~{\rm log}~ r_{KBO}$, 
where $R_0$ is the zero point of the magnitude scale, 
$r_i$ is the radius of the object, $t_1 = 2 d_{\odot} d_E$, and $t_2 =
\omega ( (1-g) \phi_1 + g \phi_2)$ \citep{bow89}.  In this last
expression, $\omega$ is the albedo, and $g$ is the slope parameter;
$\phi_1$ and $\phi_2$ are phase functions that describe the visibility
of the illuminated hemisphere of the object as a function of $\beta$.
We adopt standard values, $\omega = 0.04$ and $g = 0.15$, appropriate
for comet nuclei \citep{jew98,luu02, br04a}. The zero point $R_0$ is 
the apparent red magnitude of the Sun, $m_{R,\odot}$ = $-$27.11, 
with a correction for
the V--R color of a KBO, $R_0$ = $m_{R,\odot}$ + $\delta$(V--R)$_{KBO}$.
Observations suggest that KBOs have colors that range from roughly
$-0.1$ to 0.3 mag redder than the Sun \citep{jew01,teg03a,teg03b}. 
We treat this observation by allowing the color to vary randomly 
in this range.

For this application, we assign distances $d_{\odot}$ = 40--50 AU
and derive the number of objects in half magnitude bins for 
$R$ = 15--50.  In most models, the surface density of objects 
predicted by the model closely follows the linear relation for
log $\sigma$ with $\alpha \approx$ 0.55--0.7 and $R_0 \approx$ 
21--24 \citep[see also][]{kl99b, ken02}.
To make easier comparisons between observations and theory,
we follow \citet{ber03} and define the relative space density
as the ratio between the model and the linear surface density
relation with $\alpha$ = 0.6 and $R_0$ = 23.

Figure 10 shows the relative surface density for several KBO calculations.
For bright KBOs with $R \approx$ 22--28, the surface density
closely follows the linear relation, equation (14). This result 
is independent of the fragmentation parameters, the initial
mass at 40--50 AU, the initial size distribution, and the
amount of stirring by Neptune.  Thus, the coagulation models
provide a robust prediction for $\alpha$ at $R \approx$ 22--28
\citep[see also][]{kl99b, ken02}.

There are significant differences between the models at 
$R \lesssim$ 22 and $R \gtrsim$ 28. For $R \lesssim$ 22,
the models fall into two broad classes defined by the ratio 
of the disruption energy to the typical collision energy,
$Q_d / Q_I$. Calculations with large $Q_d/Q_I$ produce 
many large, bright KBOs.  Calculations with small $Q_d/Q_I$ 
produce few large, bright KBOs. The range in production rates 
is roughly 4 orders of magnitude. 

For fainter KBOs, the differences between models become even more 
significant.   At $R \approx$ 27--30, collisions remove weak
KBOs from the size distribution. Thus, most calculations with 
Neptune stirring exhibit a drop in the relative surface density. 
Calculations with $Q_b \lesssim 10^3$ erg g$^{-1}$, $Q_g \lesssim$ 
0.1--0.2, and no Neptune stirring also produce fewer KBOs at 
$R \approx$ 27--30.

Relative to the nominal power-law, all calculations produce 
an excess of KBOs at $R \approx$ 29--33.  For weak KBOs with
$Q_b \lesssim 10^3$ erg g$^{-1}$, the model predicts a factor
of 3 excess compared to the power-law.  For strong KBOs with
$Q_b \gtrsim 10^6$ erg g$^{-1}$, this excess grows to a factor 
of 10--30.  Stirring by Neptune has little impact on this excess.

For fainter KBOs with $R \gtrsim$ 32, stirring by Neptune is very 
important. Our models produce a 5--12 order of magnitude deficit 
of KBOs relative to the power-law surface density relation. 
Weaker KBOs produce larger deficits. Neptune stirring also
produces larger deficits. We derive the largest deficit, 12
orders of magnitude at $R \approx$ 40, for models with Neptune
stirring, $Q_b \lesssim 10^4$ erg g$^{-1}$, and $Q_g \lesssim$
0.15 erg g$^{-1}$ (Figure 10).

The derived $\sigma(m)$ yields good agreement with the observations
(Figure 11). For $R \approx$ 22--28, most models account for the
variation of relative number density with R magnitude. 
Calculations with weaker KBOs, $Q_b \lesssim 10^4$ erg g$^{-1}$,
reproduce the dip in the relative number density at $R \approx$ 
19--20.  Our ensemble of models suggests that the magnitude of
the dip depends more on stochastic phenomena than on model parameters.
The small $Q_b$ models also provide better agreement with observations
for fainter KBOs with $R \approx$ 26--28. At $R \approx$ 20--21 
and at $R \approx$ 29--30, models without Neptune stirring produce 
an excess of KBOs relative to the observations; models with Neptune 
stirring yield better agreement with the data.  Thus, models with
weak KBOs, $Q_b \lesssim 10^4$ erg g$^{-1}$, and with Neptune stirring
provide the best explanation for current observations of the shape
of the size distribution of KBOs.

In addition to the relative size distribution, the collision models
provide fair agreement with the absolute numbers of KBOs (Figure 12).
Current data suggest a total mass of $\sim$ 0.1 $M_{\oplus}$ in
KBOs at 35--50 AU \citep{luu02, ber03}.  To form KBOs by coagulation
in 10--100 Myr, collision models require an initial mass in solids 
comparable to the minimum mass solar nebula.  This result suggests
that the current mass in KBOs is $\sim$ 1\% of the initial mass in
solid material at 35--50 AU.  After 4.5 Gyr, our collision models 
have 3\%--10\% of the initial mass in 1 km and larger objects. 
Models with Neptune stirring are more efficient at removing material
from the size distribution (Figure 12).  

This result is encouraging. Once large objects form in the Kuiper
Belt, the collisional cascade can remove almost all of the leftover
planetesimals, which contain 90\% to 97\% of the initial mass
\citep[see also][]{st97a, kl99a}.
Other processes not included in our calculations will also remove 
large objects. Dynamical interactions with Neptune and other giant 
planets can remove 50\% to 80\% of the initial mass 
\citep{hol93, lev95}. Interactions with field stars can also 
remove KBOs \citep{ida00}.  Including these
processes in a more realistic collision calculation should bring
the predicted number of KBOs into better agreement with observations.
We plan to describe calculations testing the role of Neptune in 
the formation and evolution of the Kuiper Belt.

Although we cannot develop tests using data for KBO dynamical families, 
the models provide some insight into general trends of these 
observations. Because scattered KBOs and most resonant KBOs have 
had close dynamical interactions with Neptune, these objects probably 
formed closer to the Sun than classical KBOs. If Neptune stirring 
halted accretion in the Kuiper Belt, this difference in heliocentric
distance can produce an observable difference in KBO sizes. For a
formation timescale, $t \propto P/\Sigma \propto a^3$, KBOs at 
45 AU take twice as long to 
form as KBOs at 35 AU. During the late stages of runaway growth, 
this difference in formation timescales leads to a factor of 
$\sim$ 2 difference in the maximum size of a KBO.  Although the 
oligarchic growth phase erases this difference, 
significant stirring by Neptune during runaway growth might preserve 
the difference and lead to the apparent lack of large classical KBOs
(formed at $\sim$ 45 AU) relative to resonant KBOs (formed at $\sim$ 
35 AU). We plan additional numerical calculations to test this idea.

\section{CONCLUSIONS AND SUMMARY}

We have developed an analytic model for the formation of a break in 
the power law size distribution of KBOs in the outer solar system. 
For a mass in KBOs equivalent to the current mass at 40--50 AU
and for $e$ = 0.04 orbits, the model predicts a break at a radius,
$r_b \sim$ 0.1--1 km.  For a massive Kuiper Belt with $e$ = 0.2,
the break moves to $r_b \sim$ 10--100 km. These results agree with
the model of \citet{pan04}.  

In contrast to \citet{pan04}, our model predicts a smaller sensitivity 
to the bulk strength of KBOs. When the mass in KBOs is 1\% of the 
minimum mass solar nebula, $r_b$ is independent of $Q_b$ for 
$Q_b \lesssim$ $10^6$ erg g$^{-1}$. For $Q_b \gtrsim$ 
$10^6$ erg g$^{-1}$, $r_b$ declines with increasing $Q_b$. 
As the total mass in KBOs increases, the break radius is
less sensitive to $Q_b$. When the mass in KBOs is comparable
to a minimum mass solar nebula, the analytic model suggests that 
$r_b$ is roughly constant for all reasonable $Q_b$.

To test the analytic model, we used a suite of numerical simulations.
Constant eccentricity calculations, with no velocity evolution
due to gravitational interactions, confirm the analytic results.
For models with constant $e$ = 0.04--0.20, the break radius depends on
$Q_b$ and the total mass in KBOs,
\begin{equation}
r_{br} \approx \left\{ \begin{array}{l l l}
	1.0 \left ( \frac{e}{0.04} \right )^{1/2} \left ( \frac{M_{KBO}}{M_{KBO,0}} \right )^{1/2} ~ {\rm km} & \hspace{5mm} & Q_b \lesssim 10^6 ~ {\rm erg ~ g^{-1}} \\
\\
	0.1 \left ( \frac{e}{0.04} \right )^{1/2} \left ( \frac{M_{KBO}}{M_{KBO,0}} \right ) ~ {\rm km} & \hspace{5mm} & Q_b \gtrsim 10^7 ~ {\rm erg ~ g^{-1}} \\
         \end{array}
         \right .
\end{equation}
where $M_{KBO,0}$ is the mass of a minimum mass solar nebula 
extended into the Kuiper Belt, $\sim$ 10 $M_{\oplus}$ at 40--50 AU.

Simulations of complete KBO evolution, with velocity stirring,
generally require more initial mass in planetesimals to yield the 
same results. In models without Neptune formation, stirring by large 
objects with radii of 1000--3000 km yield 
\begin{equation}
r_{br} \approx \left\{ \begin{array}{l l l}
	1-3 ~ {\rm km} & \hspace{5mm} & Q_b \lesssim 10^4 ~ {\rm erg ~ g^{-1}} \\
\\
	1-3 ~ \left ( \frac{Q_b}{\rm 10^4 ~ erg ~ g^{-1}} \right )^{1/2} ~ {\rm km} & \hspace{5mm} & Q_b \gtrsim 10^4 ~ {\rm erg ~ g^{-1}} \\
         \end{array}
         \right .
\end{equation}
for models starting with a mass in solid comparable to the minimum
mass solar nebula. 
Calculations with Neptune at 30 AU allow larger break radii independent
of $Q_b$, with
\begin{equation}
r_{br} \approx \begin{array}{l l l}
3-10 ~ {\rm km} & \hspace{5mm}  & Q_b \lesssim 10^7 ~ {\rm erg ~ g^{-1}} \\
         \end{array}
\end{equation}
In both cases, models with more initial mass yield larger $r_b$. 

Comparisons between observed and predicted size distributions of KBOs
allow tests of models for KBO formation and evolution. For a broad
range of input parameters, KBO models with and without stirring by
Neptune yield good agreement with observations for $R \approx$ 21--27.
The observed surface density of brighter KBOs suggests that KBOs
have $Q_b \lesssim 10^3 - 10^4$ erg g$^{-1}$. Although stirring by
Neptune modifies the shape of the KBO size distribution for 
$R \lesssim$ 22, the observations do not discriminate clearly 
between models with and without Neptune.  Improved observational
constraints on the surface density of KBOs for $R \lesssim$ 21--22
might provide tests for the relative formation times of Neptune and
large KBOs.

Observations for $R \gtrsim$ 27 may also yield constraints on the
formation of Neptune.
Long-range stirring by Neptune is more important for the size distribution 
of fainter KBOs, $R \gtrsim$ 27.  Most models predict a small dip in the
size distribution at $R \approx$ 27--30, a peak at $R \approx$ 30--34,
and a deep trough at $R \approx$ 35--45. Because stirring by Neptune 
removes more objects with radii of 1--30 km from the KBO size distribution,
models with Neptune produce a larger dip and a deeper trough than models
without Neptune.  The depth of the small dip in models with Neptune stirring
is close to the depth observed in recent HST observations \citep{ber03}.  

Calculations with Neptune stirring yield many orders of magnitude fewer 
KBOs with $R \approx$ 32--42 than calculations without Neptune. Current
observations do not probe this magnitude range. However, ongoing and
proposed campaigns to detect small KBOs from occultations of background
stars allow tests of the models
\citep{bai76, bro97, roq00, coo03, roq03}.
For $\omega$ = 0.04, detections of
1 km KBOs provide constraints at $R \approx$ 33--35, where Neptune
stirring models predict a sharp drop in the KBO number density. 
Direct detections of smaller KBOs, with radii of $\sim$ 0.1 km,
constrain model predictions at $R \approx$ 40, where Neptune stirring
models predict 3--6 order of magnitude fewer KBOs than models without
Neptune stirring.

After 4.5 Gyr of collisional evolution, all of the numerical calculations 
predict a small residual mass in large KBOs. For $Q_b \lesssim$
$10^6$ erg g$^{-1}$, the simulations leave $f \sim$ 3\%-8\% of the initial
planetesimal mass in KBOs with radii of 1 km and larger. Models with
stirring by Neptune contain less mass in KBOs than models without
Neptune. Calculations with $Q_b \gtrsim$ $10^6$ erg g$^{-1}$ have
a larger range in $f$; models with stirring by Neptune still leave
$\sim$ 3\%--5\% of the initial mass in large KBOs. 

These results are encouraging.  Although our calculations leave more
material in large objects than the current mass in KBOs, $\sim$ 
0.5\%--1\% of a minimum mass solar nebula, other processes can 
reduce the KBO mass considerably. Formation of Neptune at 10--20 Myr,
instead of our adopted 100 Myr, probably reduces our final mass 
estimates by a factor of two.  Dynamical interactions with Neptune 
and passing stars can also remove substantial amounts of material
\citep[e.g.,][]{hol93,lev95,dun95,mal96,lev97,mor97, ida00}. 
These studies suggest that a combination of collisional grinding 
and dynamical interactions with Neptune or a passing star can 
reduce a minimum mass solar nebula to the mass observed today in 
the Kuiper Belt.  We plan to describe additional tests of these 
possibilities in future publications.

Finally, our calculations provide additional evidence that observations 
of Kuiper Belt objects probe the formation and early evolution of 
Neptune and other icy planets in the outer solar system. Better
limits on the sizes of the largest KBOs probe the timescale for
Neptune formation\footnote{The discovery of Sedna \citep{br04b}
tests models for KBO formation at 50--100 AU.}. These observations 
also constrain the bulk strength of KBOs during the 
formation epoch.  The detection of small KBOs, $r \approx$ 0.1--10 km, 
by occultations \citep[e.g. {\it TAOS};][]{mar03} or by direct 
imaging \citep[e.g., {\it OWL};][]{gil01}
yields complementary constraints.  As the observations improve,
the theoretical challenge is to combine collisional (this paper;
\cite{gol04}) and dynamical \citep[e.g.][]{mal95,gom03,lev03,qui04} 
calculations to derive robust
predictions for the formation and evolution of Uranus, Neptune, 
and smaller icy planets at heliocentric distances $\gtrsim$ 15 AU. 
Together, the calculations and the observations promise detailed 
tests of theories of planet formation.

\vskip 6ex

We acknowledge a generous allotment, $\sim$ 3000 cpu days from
February 2003 through March 2004, of computer time at the 
supercomputing center at the Jet Propulsion Laboratory through 
funding from the NASA Offices of Mission to Planet Earth, 
Aeronautics, and Space Science.  
Advice and comments from M. Geller and S. A. Stern improved 
our presentation.  We thank P. Michel and A. Morbidelli for 
extensive discussions that improved our treatment of collisional 
disruption.  We also acknowledge discussions with P. Goldreich 
and M. Holman.  The {\it NASA} {\it Astrophysics Theory Program} 
supported part of this project through grant NAG5-13278.  

\appendix

\section{APPENDIX}

\citet[][1999]{kl98} and \citet[][2002a, 2004]{kb01} describe
algorithms and tests of our multiannulus planet formation code.
Here, we describe an update to the fragmentation algorithm.

In previous calculations, we used fragmentation prescriptions
summarized by \citet{dav85} and \citet{ws93}. Both methods write
the strength $S$ of a pair of colliding objects as the sum of 
a constant bulk strength, $S_0$, and the gravitational binding 
energy, $E_g$,
\begin{equation}
S = S_0 + E_g ~ .
\end{equation}
For $E_g \propto (m_i + m_j) / r_{ij}$, the strength is
$S$ = $S_0$ + $S_1 r_{ij}^2$, where $S_1$ is a constant and
$r_{ij}$ is the radius of an object with mass $m_i + m_j$.
When the collision energy, $Q_I$, exceeds $S$, \citet{ws93} 
derive the mass lost by catastrophic disruption as the ratio 
of the impact energy to a crushing energy $Q_c$. \citet{dav85}
assume that a fixed fraction, $f_{KE}$, of the impact kinetic 
energy is transferred to ejected material and derive the
fraction of the combined mass lost to disruption. In most
cases, the \citet{dav85} algorithm yields less debris than
the \citet{ws93} algorithm.

To take advantage of recent advances in numerical simulations
of collisions \citep[e.g.,][2003]{ben99,mic01,mic02},
we now define a disruption energy $Q_d$ required to eject 50\% 
of the combined mass of two colliding bodies (equation 3 in 
the main text). For collision energy $Q_I$, the mass ejected 
in a catastrophic collision is 
$m_{ej} = 0.5 (m_1 + m_2) (Q_I/Q_d)^{\beta_e}$. For most
applications we set $\beta_e$ = 1.125 \citep[e.g.,][]{dav85}.
When $m_{ej} < 10^{-8}$ $(m_i + m_j)$, we follow \citet{ws93}
and set $m_{ej}$ = 0.

To derive the size and velocity distribution of the ejected
material, we adopt a simple procedure for all collisions. We 
define the remnant mass,
\begin{equation}
m_{rem} = m_i + m_j - m_{ej} ~ .
\end{equation}
The mass of the largest ejected body is 
\begin{equation}
m_{L,ej} = 0.2 m_{ej}
\end{equation}
We adopt a cumulative size distribution for the remaining ejected 
bodies, $n_c(m) \propto m^{-b}$, with $b$ = 0.8 \citep{doh69,wil94},
and require that the mass integrated over the size distribution 
equal $m_{ej}$. We assume
that all bodies receive the same kinetic energy per unit mass,
given by the initial relative velocities of the two bodies,
\begin{equation}
V_{ij}^2 = h_i^2 + v_i^2 + h_j^2 + v_j^2 ~ ,
\end{equation}
where $h$ and $v$ are the horizontal and vertical components of the 
velocity dispersion relative to a circular orbit \citep[e.g.][]{kl98}.
We derive mass-weighted $h_i^{\prime}$ and $v_i^{\prime}$ for
the combined object and the debris, $V_{ij}^2$ = $(h_i^{\prime})^2$ 
+ $(v_i^{\prime})^2$.

This procedure, which we apply to cratering and disruptive collisions,
is computationally efficient and maintains the spirit of recent analytic 
models and numerical simulations. Comparisons with our previous results
using the \citet{dav85} and \citet{ws93} algorithms suggest that the
new algorithm yields intermediate `mass loss rates' for $\beta_g$ = 2
and $Q_b \sim 10^2$--$10^6$ erg g$^{-1}$.  Calculations with $\beta_g$
$\sim$ 1.2--1.5 \citep{ben99} yield larger mass loss but do not change
the results significantly.  When the bulk strength depends on the particle
radius, the size distribution for small objects with $r_i \lesssim$
0.1 km depends on the exponent $\beta_b$ of the bulk component of
the strength. We plan to report on the details of these differences
in future papers on the formation of KBOs and terrestrial planets.

\clearpage



\begin{figure}
\plotone{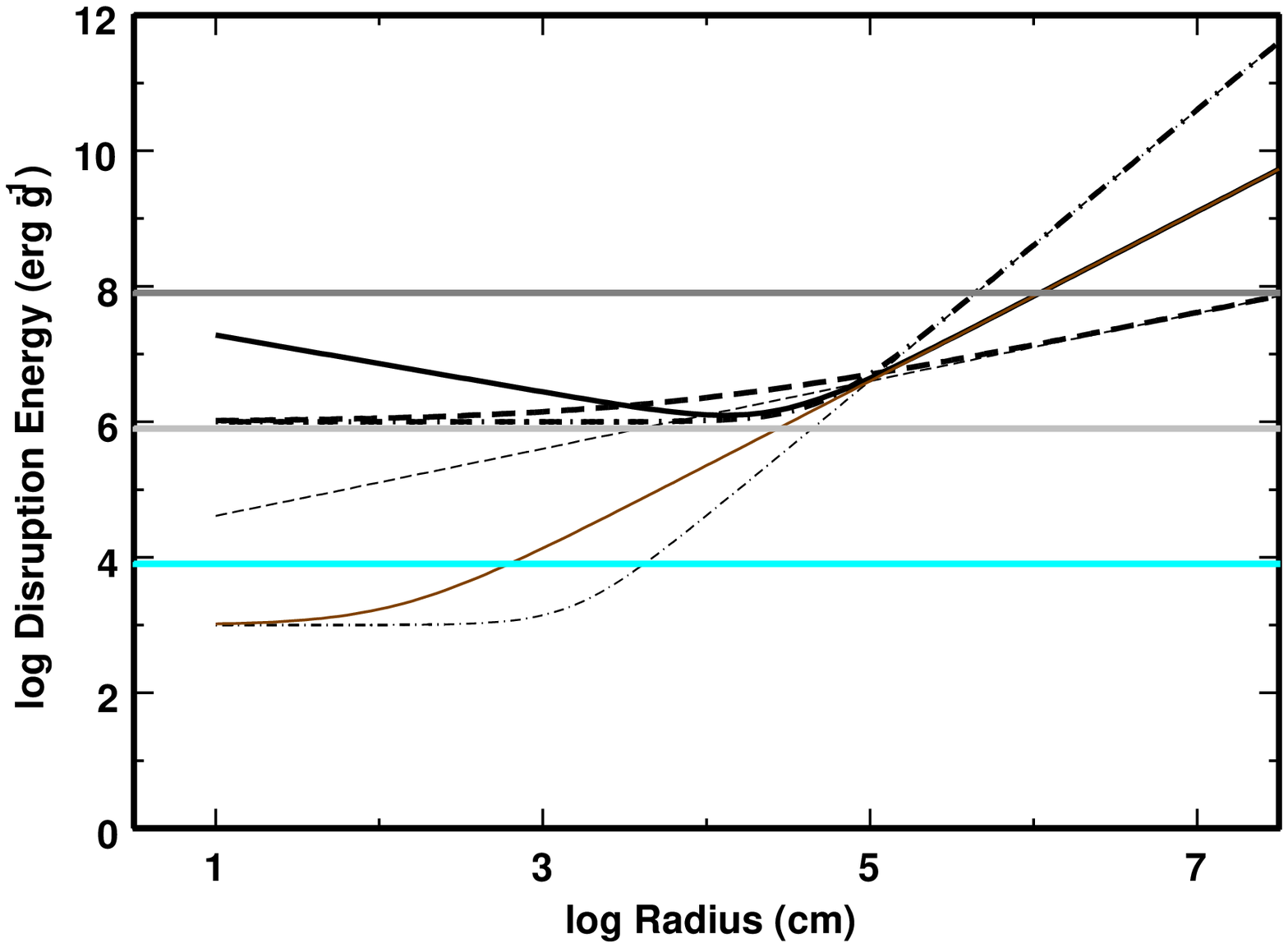}
\caption
{Comparison of catastrophic disruption energy for models 
described by equation (3) with $\rho$ = 1.5 g cm$^{-3}$.
Solid curve: 
$Q_b = 1.6 \times 10^7$ erg g$^{-1}$,
$Q_g = 1.5 $ erg g$^{-1}$,
$\beta_b$ = $-0.42$,
$\beta_g$ = 1.25 \citep{ben99}.
Dot-dashed curves: 
$Q_g = 2.67 \times 10^{-4} $ erg g$^{-1}$,
$\beta_b$ = $0$,
$\beta_g$ = 2 \citep{dav85}.
Dashed curves: 
$Q_g = 8.4 \times 10^{3} $ erg g$^{-1}$,
$\beta_b$ = $0$,
$\beta_g$ = 0.5 \citep{dav85}.
The light lines have 
$Q_b = 10^3$ erg g$^{-1}$;
the heavy lines have
$Q_b = 10^6$ erg g$^{-1}$.
The horizontal lines indicate the collision energy for
KBOs with $e$ = 0.001 (light grey),
KBOs with $e$ = 0.01 (medium grey),
KBOs with $e$ = 0.1 (heavy grey).}
\end{figure}
\clearpage

\begin{figure}
\plotone{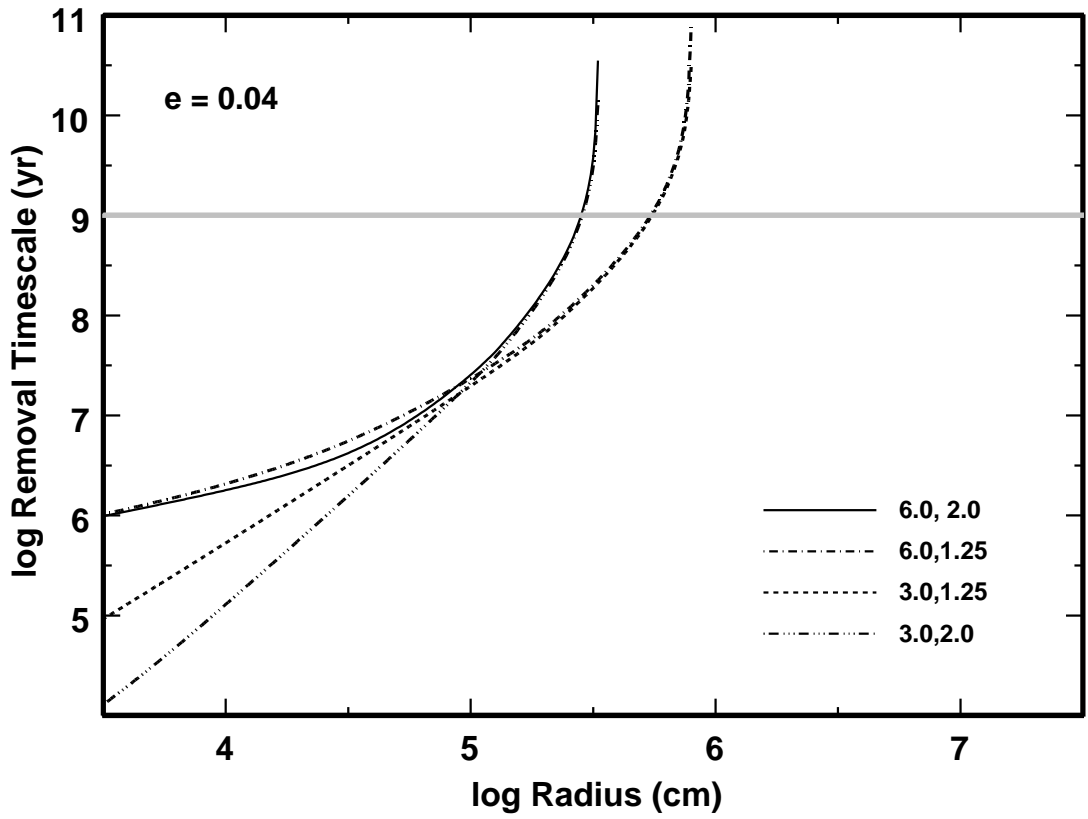} 
\caption
{Removal timescale as a function of size for KBO collisions with
$e$ = 0.04. The legend lists log $Q_b$ and $\beta_g$ for each curve.}
\end{figure}

\begin{figure}
\plotone{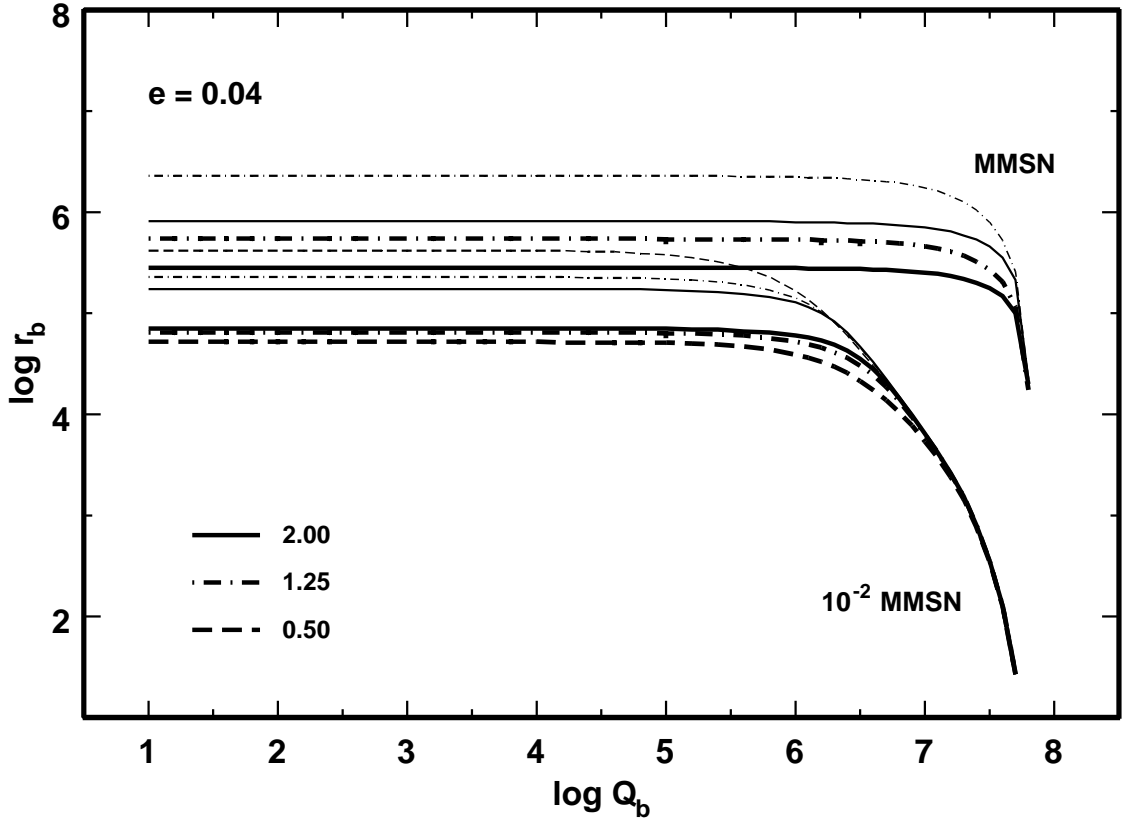} 
\caption
{Break radius, $r_b$, as a function of the bulk strength,
log $Q_b$, for collisions with $e$ = 0.04. The legend
lists $\beta_g$ for each curve. Heavy lines have 
$C_g = 2.25 \times 10^{5(1.25 - \beta_g)}$ erg g$^{-1}$;
light lines have
$C_g = 0.225 \times 10^{5(1.25 - \beta_g)}$ erg g$^{-1}$.
The upper set of curves plots results for a minimum mass
solar nebula; the lower set plots results for models with
1\% of the mass in a minimum mass solar nebula.}
\end{figure}

\begin{figure}
\plotone{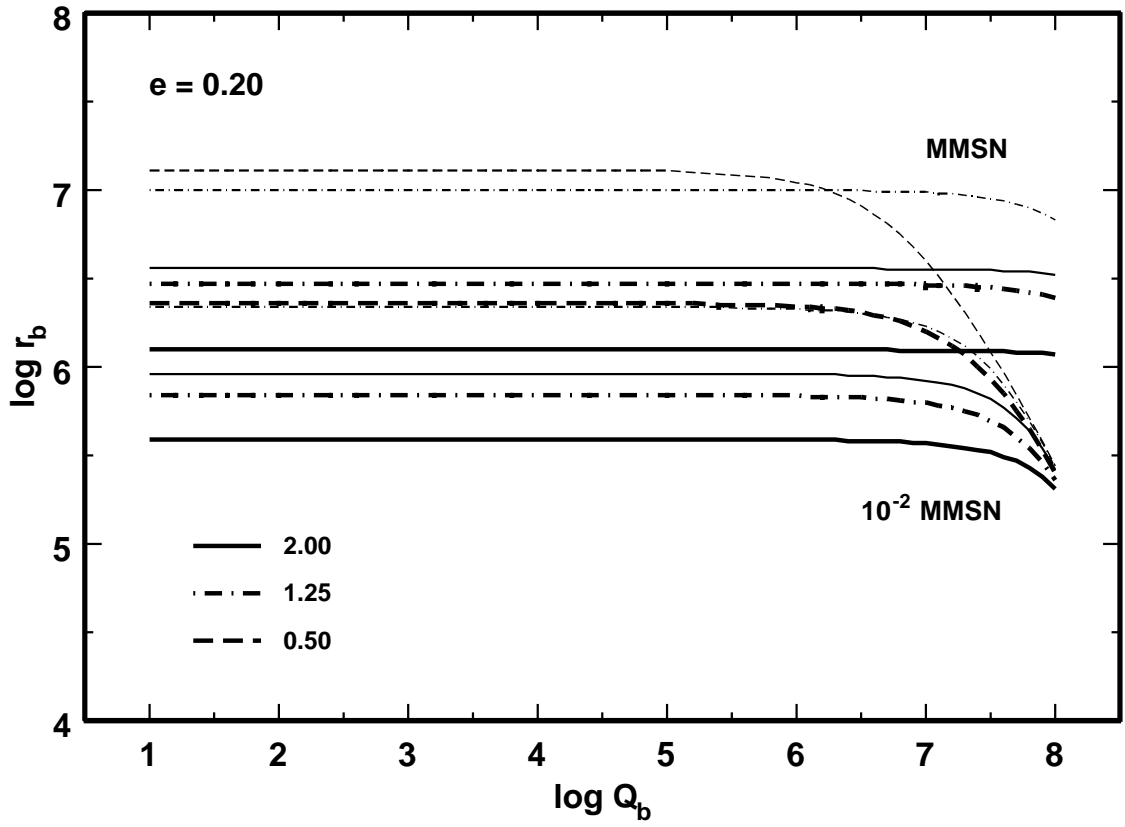} 
\caption
{As in Figure 4, for $e$ = 0.2.}
\end{figure}

\begin{figure}
\plotone{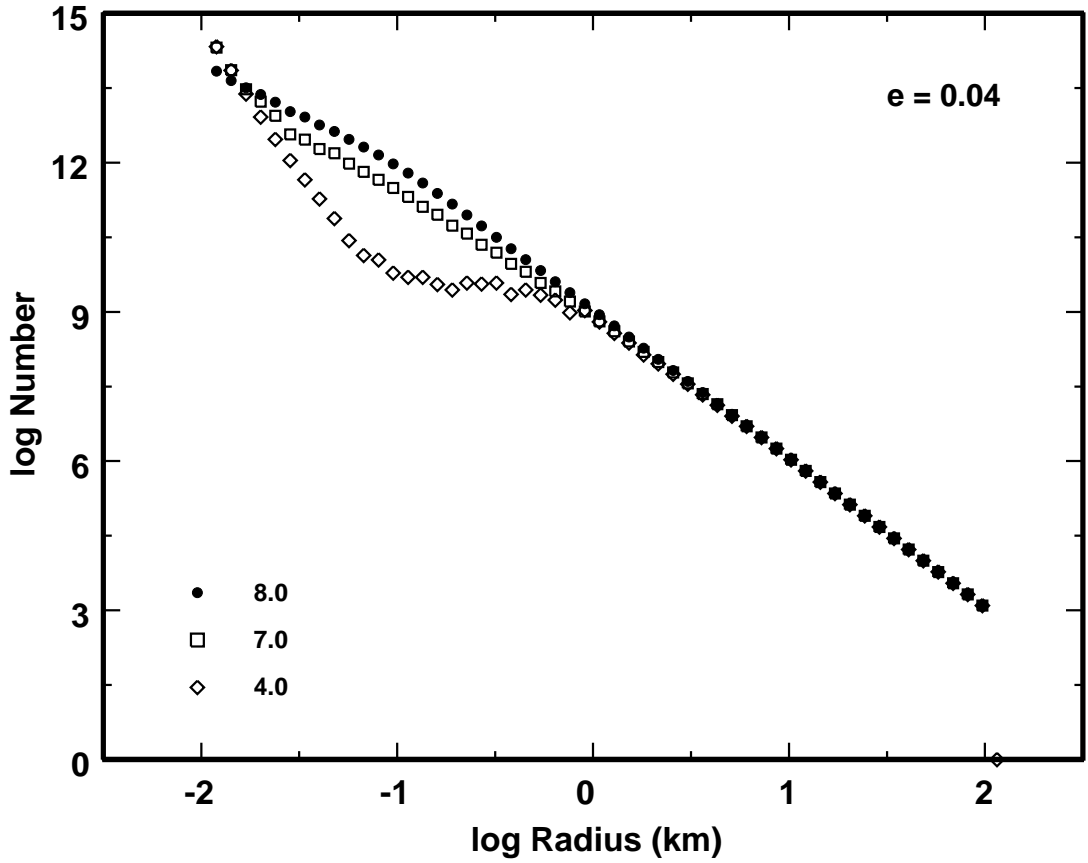} 
\caption
{Size distributions at 4.5 Gyr for numerical models with
constant $e$ = 0.04. The legend lists log $Q_b$ for each 
model.}
\end{figure}

\begin{figure}
\plotone{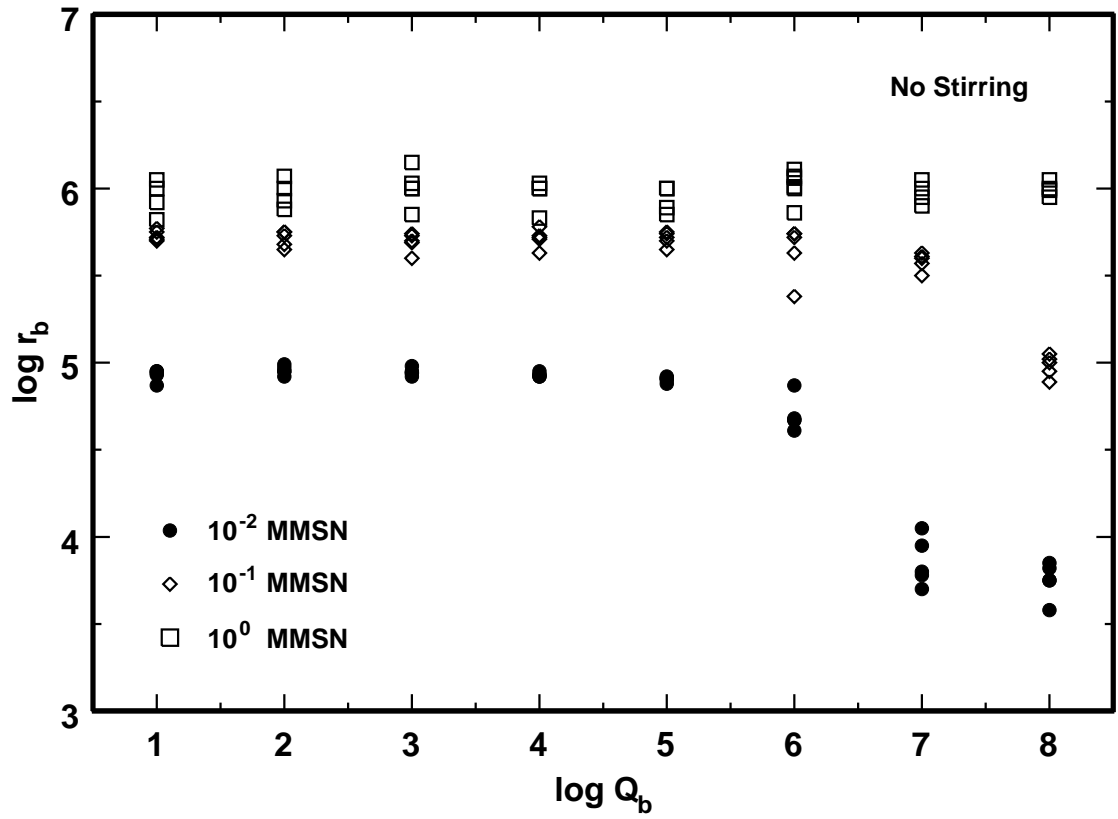} 
\caption
{Variation of $r_b$ with $Q_b$ for numerical calculations
at 40-47 AU with constant $e$=0.04. The legend indicates 
the initial mass in solids for each set of calculations.}
\end{figure}

\begin{figure}
\plotone{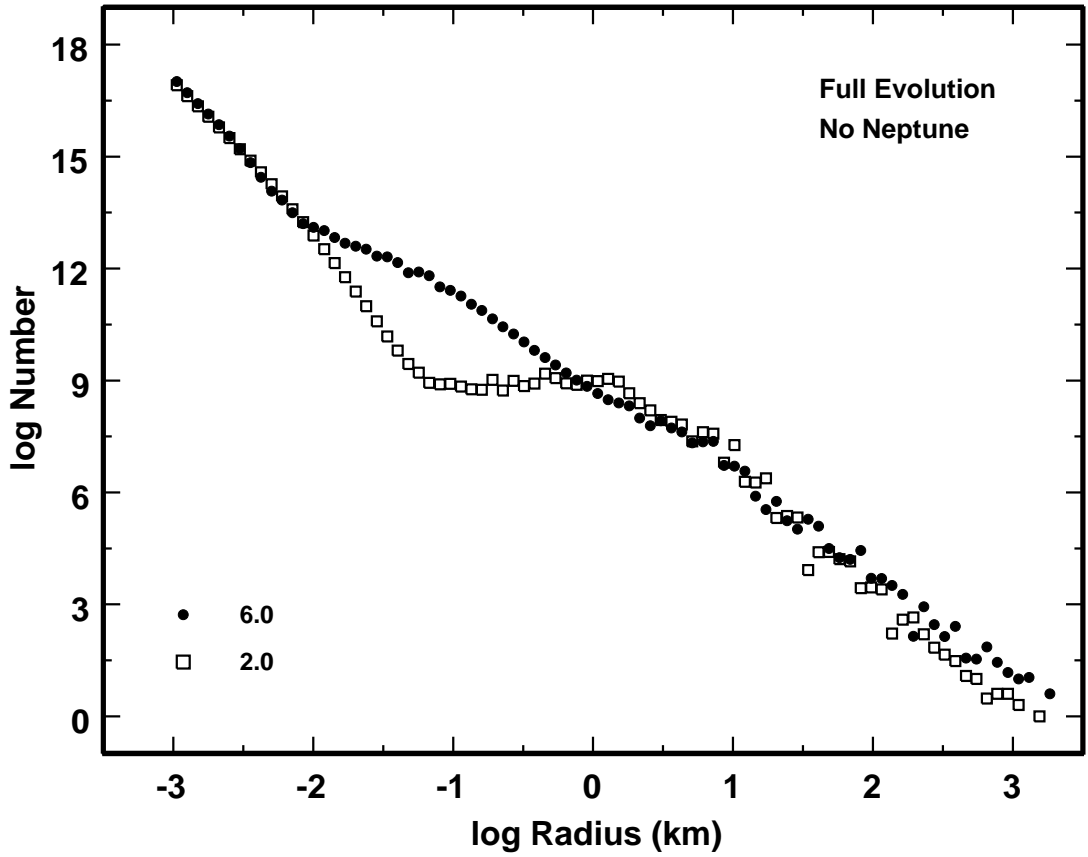} 
\caption
{Size distributions at 4.5 Gyr for complete numerical models 
of KBO evolution at 40--47 AU.  The legend lists log $Q_b$ 
for each model.}
\end{figure}

\begin{figure}
\plotone{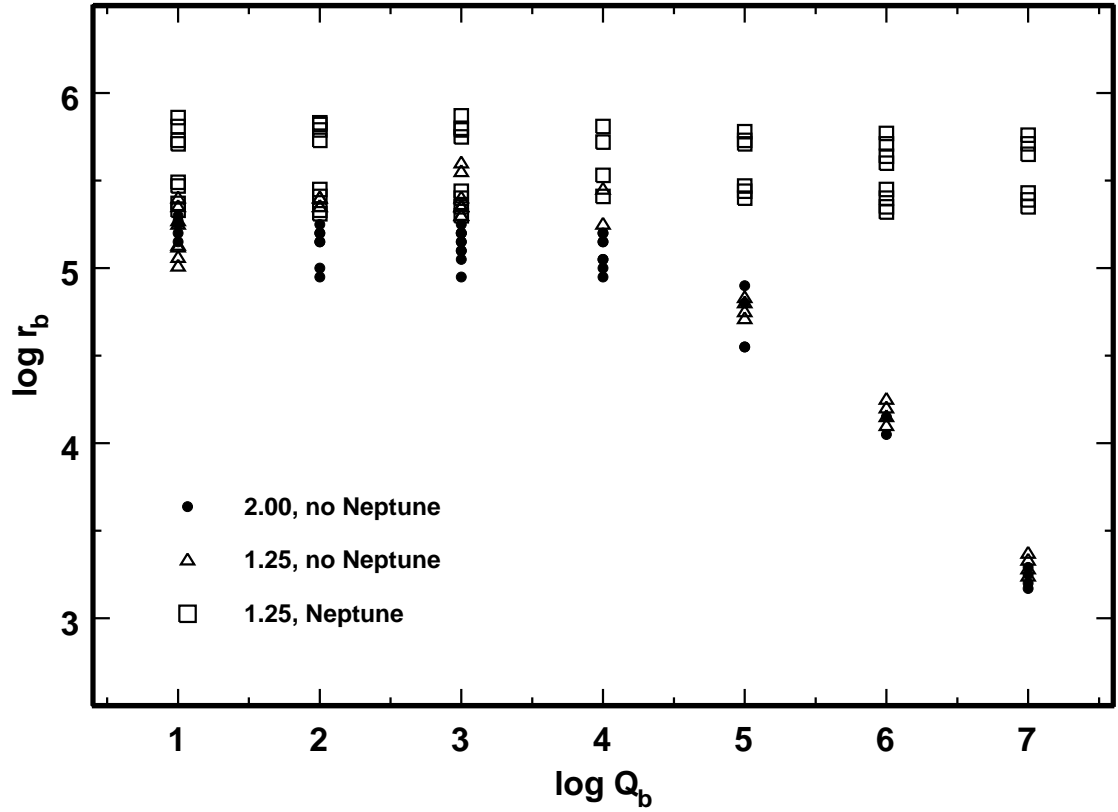} 
\caption
{Variation of $r_b$ with $Q_b$ for complete numerical 
calculations of KBO evolution at 40-47 AU. All calculation
begin with a mass in solids equivalent ot the minimum mass
solar nebula. The legend indicates log $Q_b$ for each set
of models. Calculations with Neptune use a simple model
for the growth of Neptune at 30 AU.}
\end{figure}

\begin{figure}
\plotone{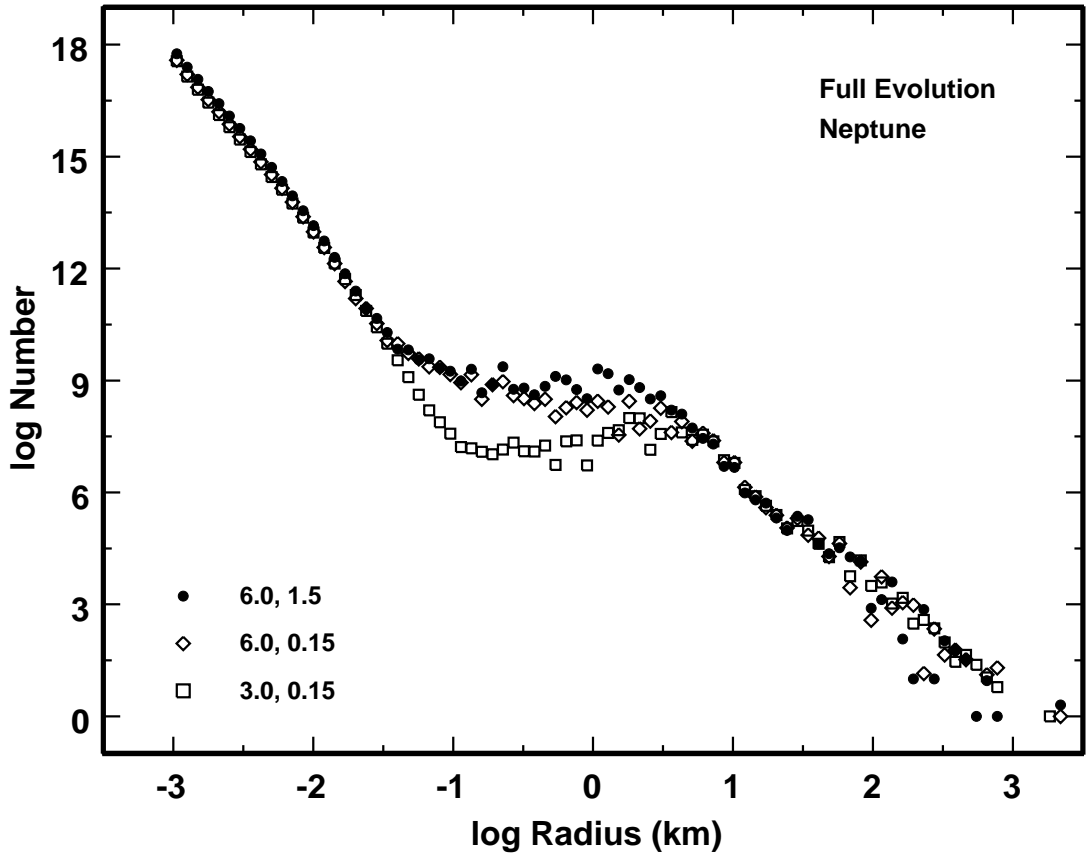} 
\caption
{As in Figure 7, for models with Neptune at 30 AU.}
\end{figure}

\begin{figure}
\plotone{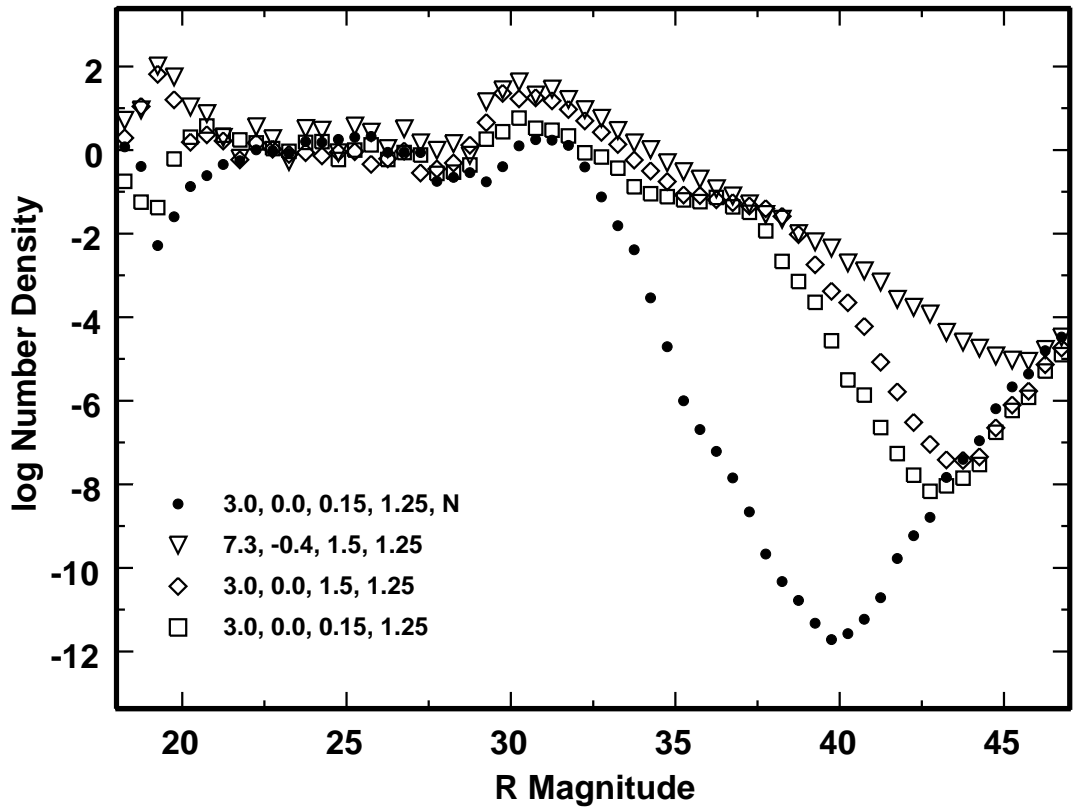} 
\caption
{KBO luminosity functions derived from the planet formation model.  
The legend indicates fragmentation models for each model;
the model with `N' has Neptune stirring.}
\end{figure}

\begin{figure}
\plotone{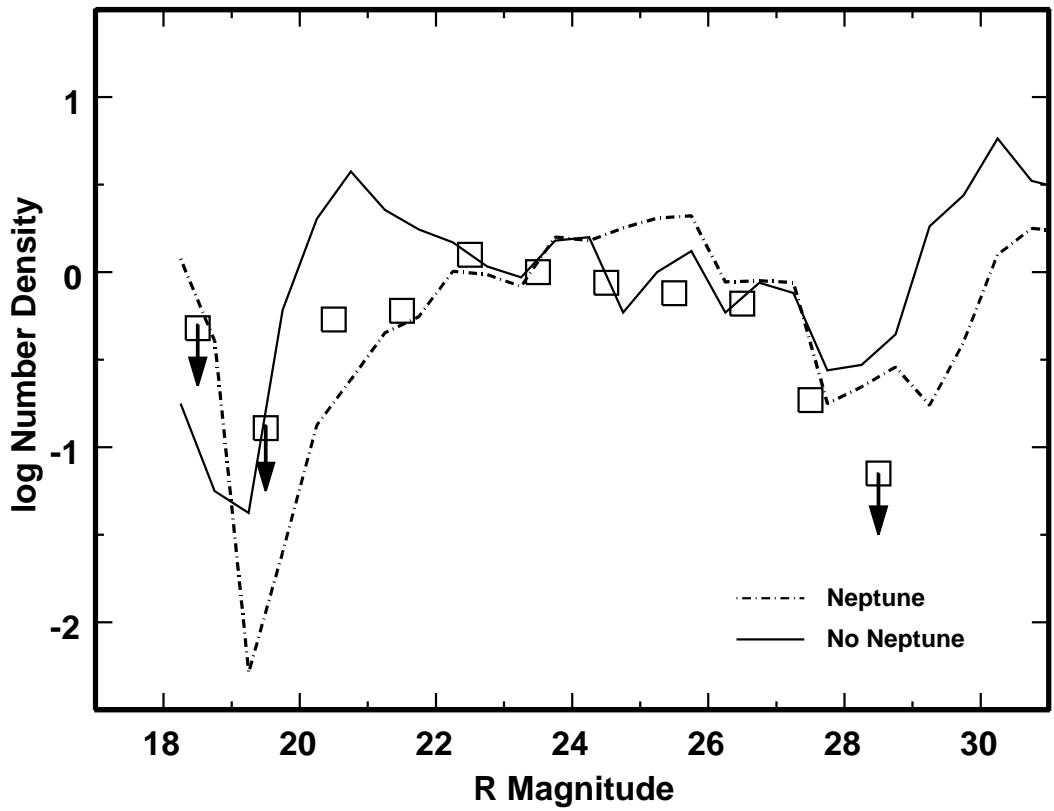} 
\caption
{As in Figure 10, with observations of KBOs added for comparison
\citep{ber03}.}
\end{figure}

\begin{figure}
\plotone{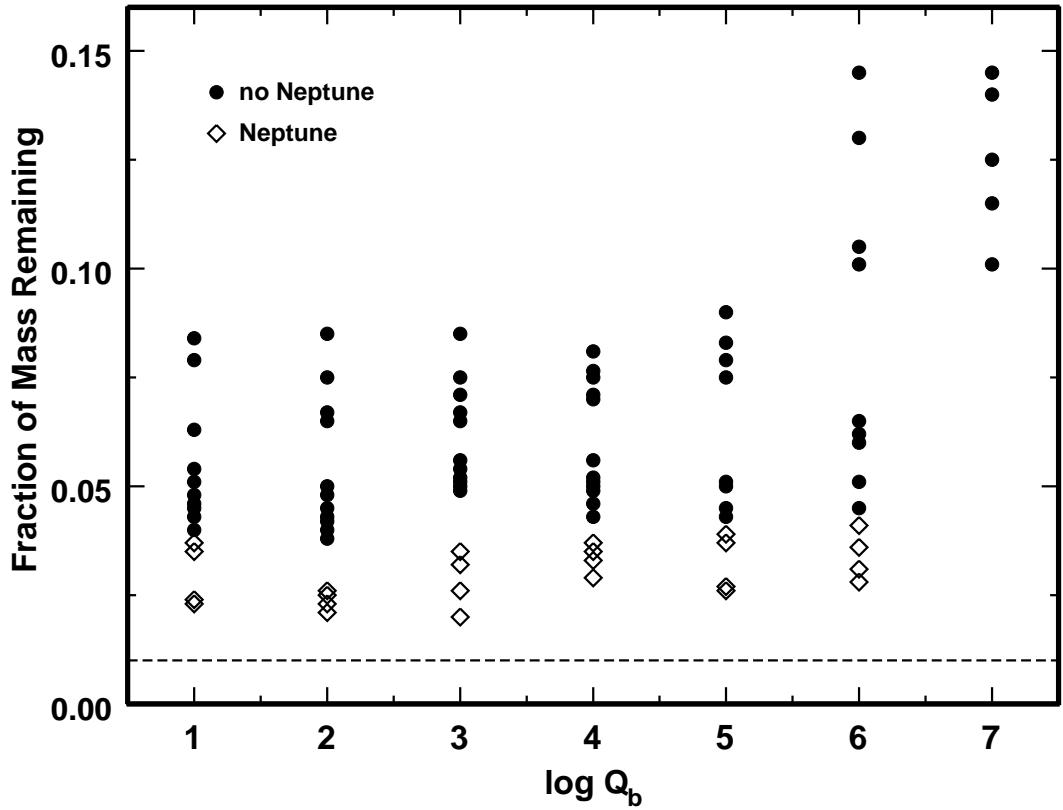} 
\caption
{Fraction of initial mass remaining in complete KBO models 
at 4.5 Gyr. Filled circles: models without Neptune.
Open circles: models with Neptune at 30 AU.
At the bottom of the plot, the horizontal dashed line indicates
the ratio of the current mass in the Kuiper Belt to the
mass in solids of a minimum mass solar nebula.}
\end{figure}

\clearpage


\begin{thebibliography}{99}
\vskip 4ex

\baselineskip=12pt
\parskip=0pt

\bibitem[Adachi et al. (1976)]{ada76} Adachi, I., Hayashi, C., \& Nakazawa, K.
1976, Progress of Theoretical Physics 56, 1756

\bibitem[Allen et al. (2001)]{all01} Allen, R. L., Bernstein, G. M.,
\& Malhotra, R. 2001, ApJ, 549, L241

\bibitem[Artymowicz (1997)]{art97}Artymowicz, P. 1997, ARE\&PS, 25, 175

\bibitem[Asphaug \& Benz (1996)]{asp96} Asphaug, E., \& Benz, W. 1996,
Icarus, 121, 225

\bibitem [Backman \& Paresce (1993)]{bac93}
Backman, D. E., \& Paresce, F. 1993, in
Protostars and Planets III, ed. E. H. Levy \& J. I. Lunine
(Tucson: Univ of Arizona), p. 1253

\bibitem [Backman,~Dasgupta, \& Stencel (1995)]{bac95}
Backman, D. E., Dasgupta, A., \& Stencel, R. E. 1995, ApJ, 450, L35

\bibitem[Bailey (1976)]{bai76} Bailey, M. 1976, Nature, 259, 290

\bibitem[Benz \& Asphaug (1999)]{ben99} Benz, W., \& Asphaug, E. 1999,
Icarus, 142, 5

\bibitem[Bernstein et al. (2003)]{ber03} Bernstein, G. M., Trilling, D. E., 
Allen, R. L., Brown, M. E., Holman, M., \& Malhotra, R. 2003, ApJ, submitted

\bibitem[Bowell et al. (1989)]{bow89} Bowell, E., Hapke, B., Domingue, D.,
Lumme, K., Peltoniemi, J., and Harris, A. W. 1989, in {\it Asteroids II,}
edited by R. P. Binzel, T. Gehrels, and M. S. Matthews, Tucson, Univ. of
Arizona Press, p. 524

\bibitem[Brown (2001)]{bro01} Brown, M. E. 2001, AJ, 121, 2804

\bibitem[Brown \& Trujillo (2004)]{br04a} Brown, M. E., \& Trujillo, C. 
2004, AJ, 127, 2413

\bibitem[Brown,~Trujillo, \& Rabinowitz (2004)]{br04b} Brown, M. E.,
Trujillo, C., \& Rabinowitz, D. 2004, ApJL, submitted

\bibitem[Brown \& Webster (1997)]{bro97} Brown, M. J. I., \& Webster, 
R. L. 1997, MNRAS, 289, 783

\bibitem[Burns,~Lamy, \& Soter (1979)]{bur79} Burns, J. A., Lamy, P. L.,
\& Soter, S. 1979, Icarus, 40, 1

\bibitem[Cochran et al. (1998)]{coc98} Cochran, A. L., Levison, H. F.,
Tamblyn, P., Stern, S. A., \& Duncan, M. 1998, ApJ, 455, L89

\bibitem[Cooray \& Farmer (2003)]{coo03} Cooray, A., \& Farmer, A. J. 2003,
ApJ, 587, L125

\bibitem[Davis et al. (1985)]{dav85} Davis, D. R., Chapman, C. R.,
Weidenschilling, S. J., \& Greenberg, R. 1985, Icarus, 62, 30

\bibitem[Davis \& Farinella (1997)]{dav97} Davis, D. R., \&
Farinella, P. 1997, Icarus, 125, 50

\bibitem[Dohnanyi (1969)]{doh69} Dohnanyi, J. W. 1969, J. Geophys. Res., 74, 2531

\bibitem[Duncan \& Levison (1997)]{dun97}Duncan, M. J., \& Levison, H. F. 
1997, Science, 276, 1670

\bibitem[Duncan et al. (1995)]{dun95} Duncan, M. J., Levison, H. F.,
\& Budd, S. M. 1995,  AJ, 110, 3073

\bibitem[Duncan et al. (1988)]{dun88} Duncan, M., Quinn, T., \&
Tremaine, S. 1988,  ApJL, 328, L69

\bibitem[Gilmozzi,~Rierckx, \& Monnet (2001)]{gil01}
Gilmozzi, R., Dierickx, P., \& Monnet, G. 2001, in
{\it Quasars, AGNs and Related Research Across 2000,}
A Conference on the occasion of L. Woltjer's 70th birthday, 
edited by G. Setti and J.-P. Swings, Springer, p. 184.

\bibitem[Gladman \& Kavelaars (1997)]{gla97} Gladman, B., \& Kavelaars, J. J.
1997, A\&A, 317, L35

\bibitem[Gladman et al. (1998)]{gla98} Gladman, B., Kavelaars, J. J.,
Nicholson, P. D., Loredo, T. J., \& Burns, J. A.  1998, AJ, 116, 2042

\bibitem[Gladman et al. (2001)]{gla01} Gladman, B., Kavelaars, J. J.,
Petit, J.-M., Morbidelli, A., Holman, M., \& Loredo, T. J. 2001, AJ, 122, 1051

\bibitem[Goldreich,~Lithwick, \& Sari (2004)]{gol04} Goldreich, P., 
Lithwick, Y., \& Sari, R. 2004, ARA\&A, in press (astro-ph/0405215)

\bibitem[Gomes (2003)] {gom03} Gomes, R. S. 2003, Icarus, 161, 404

\bibitem[Hahn (2003)]{hahn03} Hahn, J. M. 2003, ApJ, 595, 531

\bibitem[Hayashi (1981)]{hay81} Hayashi, C. 1981, Prog Theor Phys Suppl, 70, 35

\bibitem[Holman \& Wisdom (1993)]{hol93}
Holman, M. J., \& Wisdom, J. 1993,  AJ, 105, 1987

\bibitem[Holsapple (1994)]{hls94} Holsapple, K. A. 1994,
Planet. Space Sci., 42, 1067
 
\bibitem[Housen \& Holsapple (1990)]{hou90} Housen, K., \& Holsapple, K. 
1990, Icarus, 84, 226

\bibitem[Housen \& Holsapple (1999)]{hou99} Housen, K., \& Holsapple, K. 
1999, Icarus, 142, 21

\bibitem[Ida et al. (2000)]{ida00} Ida, S., Larwood, J., \& Burkert, A.
2000, ApJ, 528, 351

\bibitem[Ip \& Fern\'andez (1997)]{ip97} Ip, W.-H., \& Fern\'andez, J. A.
1997, A\&A, 324, 778

\bibitem[Jewitt \& Luu (1993)]{jew93} Jewitt, David C., \& Luu, Jane X.,
1993, Nature, 362, 730

\bibitem[Jewitt \& Luu (2001)]{jew01} Jewitt, David C., \& Luu, Jane X.,
2001, AJ, 122, 2099

\bibitem[Jewitt et al. (1998)]{jew98} Jewitt, D., Luu, J. X., \& Trujillo, C.
1998, AJ, 115, 2125

\bibitem[Jura (2004)]{jur04} Jura, M. 2004, ApJ, 603, 729

\bibitem[Kenyon (2002)]{ken02} Kenyon, S. J., 2002, PASP, 114, 265

\bibitem[Kenyon \& Bromley (2001)]{kb01} Kenyon, S. J., \& Bromley, B. C.
2001, AJ, 121, 538

\bibitem[Kenyon \& Bromley (2002)]{kb02} Kenyon, S. J., \& Bromley, B. C., 
2002, AJ, 123, 1757

\bibitem[Kenyon \& Bromley (2004)]{kb04} Kenyon, S. J., \& Bromley, B. C., 
2004, AJ, 127, 513

\bibitem[Kenyon \& Luu (1998)]{kl98} Kenyon, S. J., \& Luu, J. X. 1998,
AJ, 115, 2136
 
\bibitem[Kenyon \& Luu (1999a)]{kl99a} Kenyon, S. J., \& Luu, J. X.  1999a, 
AJ, 118, 1101

\bibitem[Kenyon \& Luu (1999b)]{kl99b} Kenyon, S. J., \& Luu, J. X.  1999b, 
ApJ, 526, 465

\bibitem[Kenyon \& Windhorst (2001)]{kw01} Kenyon, S. J., \& Windhorst, R.
2001, ApJ, 547, L69

\bibitem[Kuchner,~Brown, \& Holman (2002)]{kuc02} Kuchner, M J., Brown, M. E.,
\& Holman, M. 2002, AJ, 124, 1221

\bibitem[Lagrange et al. (2000)]{lag00} Lagrange, A.-M., Backman, D.,
\& Artymowicz, P. 2000, in Protostars \& Planets IV, ed.  V. Mannings,
A. P. Boss, \& S. S. Russell (Tucson: Univ. of Arizona), p. 639

\bibitem[Larsen et al. (2001)]{lar01} Larsen, J. A., et al. 2001, AJ, 121, 562

\bibitem[Levison \& Duncan (1990)]{lev90} Levison, H. F., \& Duncan, M. J.
1990, AJ, 100, 1669

\bibitem[Levison \& Duncan (1993)]{lev93} Levison, H. F., \&
Duncan, M. J. 1993,  ApJ, 406, L35

\bibitem[Levison \& Duncan (1994)]{lev94} Levison, H. F., \&
Duncan, M. J. 1994,  Icarus, 108, 18

\bibitem[Levison \& Duncan (1997)]{lev97} Levison, H. F., \& Duncan, M. J.
1997, Icarus, 127, 13

\bibitem[Levison \& Stern (2001)]{lev01}Levison, H. F., \& Stern, S. A.,
2001, AJ, 121, 1730

\bibitem[Levison \& Morbidelli (2003)]{lev03} Levison, H. F., \&
Morbidelli, A. 2003, Nature, 426, 419

\bibitem[Levison,~Lissauer, \& Duncan (1998)]{lev98} Levison, H. F.,
Lissauer, J. J., \& Duncan, M. J.  1998, AJ, 116, 1998

\bibitem[Levison \& Stern (1995)]{lev95} Levison, H. F., \& Stern, S. A.
1995, Lunar Planet Sci. Conf. XXVI, 841

\bibitem[Lissauer (1987)]{lis87} Lissauer, J. J. 1987, Icarus, 69, 249

\bibitem[Lissauer (1993)]{lis93} Lissauer, J. J. 1993, ARA\&A, 31, 129

\bibitem[Love \& Ahrens (1996)]{lov96} Love, S. G., \& Ahrens, T. J. 1996,
Icarus, 124, 141

\bibitem[Luu \& Jewitt (2002)]{luu02}
Luu, J. X., \& Jewitt, D. 2002,  ARA\&A, 40, 63

\bibitem[Luu et al (1997)]{luu97} Luu, J. X., Marsden, B., Jewitt, D.,
Trujillo, C. A., Hergenother, C. W., Chen, J., \& Offutt, W. B. 1997,
Nature, 387, 573

\bibitem[Malhotra (1995)] {mal95} Malhotra, R. 1995,
AJ, 110, 420

\bibitem[Malhotra (1996)]{mal96} Malhotra, R. 1996,  AJ, 111, 504

\bibitem[Marshall et al. (2003)]{mar03} Marshall, S., et al. 2003,
Bull. AAS, 202, 3806

\bibitem[Michel,~Benz, \& Richardson (2003)]{mic03}
Michel, P., Benz, W., \& Richardson, D. C. 2003, Nature, 421, 608

\bibitem[Michel et al. (2001)]{mic01}
Michel, P., Benz, W., Tanga, P., \& Richardson, D. C. 2001, Science, 294, 1696

\bibitem[Michel et al. (2002)]{mic02}
Michel, P., Tanga, P., Benz, W., \& Richardson, D. C. 2002, Icarus, 160, 10

\bibitem[Morbidelli,~Jacob, \& Petit, J.-M. (2002)]{mor02} Morbidelli, A., 
Jacob, C., \& Petit, J.-M. 2002, Icarus, 157, 241

\bibitem[Morbidelli \& Levison (2003)]{mor03} Morbidelli, A., \&
Levison, H. F. 2003, Comptes Rendus Phys., 4, 809

\bibitem[Morbidelli \& Valsecchi (1997)]{mor97} Morbidelli, A., \&
Valsecchi, G. B. 1997, Icarus, 128, 464

\bibitem[Ohtsuki,~Stewart, \& Ida (2002)]{oht02} Ohtsuki, K., Stewart, 
G. R., \& Ida, S. 2002, Icarus, 155, 436

\bibitem[Pan \& Sari (2004)]{pan04} Pan, M., \& Sari, R. 2004, 
Icarus, submitted (astro-ph/0402138)

\bibitem[Pollack et al. (1996)]{pol96} Pollack, J. B., Hubickyj, O.,
Bodenheimer, P., Lissauer, J. J., Podolak, M., \& Greenzweig, Y. 1996,
Icarus, 124, 62

\bibitem[Quillen,~Trilling, \& Blackman (2004)]{qui04}
Quillen, A., Trilling, D. E., \& Blackman, E. G. 2004, AJ,
submitted (astro-ph/0401372)

\bibitem[Roques \& Moncuquet (2000)]{roq00} Roques, F., \& Moncuquet, M. 
2000, Icarus, 147, 530

\bibitem[Roques et al. (2003)]{roq03} Roques, F., Moncuquet, M., 
Lavilloni\'ere, N., Auvergne, M., Chevreton, M., Colas, F., 
\& Lecacheux, J. 2003, ApJ, 594, L63

\bibitem[Stern (1995)]{ste95} Stern, S. A. 1995, AJ, 110, 856

\bibitem[Stern (1996a)]{ste96a} Stern, S. A. 1996a, AJ, 112, 1203

\bibitem[Stern (1996b)]{ste96b} Stern, S. A. 1996b, A\&A, 310, 999

\bibitem[Stern,~Bottke \& Levison (2003)]{ste03} Stern, S. A., 
Bottke, W. F., \& Levison, H. F. 2003, AJ, 125, 902

\bibitem[Stern \& Colwell (1997a)]{st97a} Stern, S. A., \&
Colwell, J. E. 1997a,  AJ, 114, 841

\bibitem[Stern \& Colwell (1997b)]{st97b} Stern, S. A., \&
Colwell, J. E. 1997b,  ApJ, 490, 879

\bibitem[Takeuchi \& Artymowicz (2001)]{tak01} Takeuchi, T., \&
Artymowicz, P. 2001, ApJ, 557, 990

\bibitem[Tegler \& Romanishin (2003)]{teg03a}
Tegler, S. C., \& Romanishin, W. 2003, Icarus, 161, 181

\bibitem[Tegler,~Romanishin, \& Consolmagno (2003)]{teg03b}
Tegler, S. C., Romanishin, W., \& Consolmagno, S. J., G. J., 
2003, ApJ, 599, 49

\bibitem[Teplitz et al. (1999)]{tep99} Teplitz, V. I., Stern, S. A.,
Anderson, J. D., Rosenbaum, D., Scalise, R. I., \& Wentzler, P. 1999,
ApJ, 516, 425

\bibitem[Trujillo,~Jewitt, \& Luu (2001)]{tru01} Trujillo, C. A., 
Jewitt, D. C., \& Luu, J. X. 2001, AJ, 122, 457

\bibitem[Tombaugh (1946)]{tom46} Tombaugh, C. 1946, 
Leaflets of the Astr. Soc. Pac., 5, 73

\bibitem[Weidenschilling (1977)]{wei77} Weidenschilling, S. J. 1977,
Ap Sp Sci, 51, 153

\bibitem[Weidenschilling (1989)]{wei89} Weidenschilling, S. J. 1989,
Icarus, 80, 179

\bibitem[Weidenschilling \& Cuzzi (1993)]{wei93} Weidenschilling, S. J.,
\& Cuzzi, J. N., 1993, in Protostars and Planets III,
ed. E. H. Levy \& J. I. Lunine (Tucson: Univ of Arizona), p. 1031

\bibitem[Weidenschilling et al. (1997)]{wei97} Weidenschilling, S. J.,
Spaute, D., Davis, D. R., Marzari, F., \& Ohtsuki, K. 1997, Icarus, 
128, 429

\bibitem[Wetherill \& Stewart (1993)]{ws93} Wetherill, G. W., \& 
Stewart, G. R.  1993, Icarus, 106, 190

\bibitem[Williams \& Wetherill (1994)]{wil94} Williams, D. R., \&
Wetherill, G. W. 1994, Icarus, 107, 117

\end{thebibliography}
\end{document}